\documentclass[aps,prd,floatfix,nofootinbib,superscriptaddress,twocolumn,showkeys]{revtex4-2}

\usepackage{amsmath}

\usepackage{amssymb}
\allowdisplaybreaks

\usepackage{caption}
\usepackage{subcaption}
\usepackage{graphicx}

\usepackage{makeidx}
\usepackage{color}
\usepackage{amsfonts}
\usepackage[usenames,dvipsnames]{pstricks}

\usepackage{epsfig}
\usepackage{pst-grad} 
\usepackage{mathtools}
\usepackage[colorlinks,hyperindex]{hyperref}
\usepackage{array}

\hypersetup
{	colorlinks,%
citecolor=black,%
linkcolor=black,%
urlcolor=black,%
}

\usepackage{allpurpose}

\usepackage{flushend}

\newcommand\nn{{\nonumber}}

\begin{document}

\title{Effect of electric interaction on the deflection and gravitational lensing in the strong field limit}

\author{Shangjie Zhou}
\address{School of Physics and Technology, Wuhan University, Wuhan, 430072, China}

\author{Muchun Chen}
\address{School of Physics and Technology, Wuhan University, Wuhan, 430072, China}

\author{Junji Jia}
\email[Corresponding author:~]{junjijia@whu.edu.cn}
\address{Center for Astrophysics \& MOE Key Laboratory of Artificial Micro- and Nano-structures, School of Physics and Technology, Wuhan University, Wuhan, 430072, China}

\date{\today}

\begin{abstract}
	The deflection angle $\Delta\phi$ of charged signals in general charged spacetime in the strong field limit is analyzed in this work using a perturbative method generalized from the neutral signal case. The solved $\Delta\phi$ naturally contains the finite distance effect and takes a quasi-power series form with a logarithmic divergence at the leading order. The coefficients of the series  contain both the gravitational and electric contributions. 
	Using the Reissner-Nordstrom spacetime as an example, we found that an electric repulsion (or attraction) tends to decrease (or increase) the critical impact parameter $b_c$. If the repulsion is strong enough, then $b_c$
	can shrink to zero and the critical particle sphere $r_{0c}$ will disappear. 
	These results are applied to the gravitational lensing of charge signal, from which we solved the image positions, their magnifications and time delays. It is found that in general, the electric repulsion (or attraction) will decrease (or increase) the image apparent angles, the black hole shadow size as well as their magnifications but increase (or decrease) the time delay.
	
\end{abstract}


\maketitle

\section{Introduction}

Deflection of light in gravitational field was one of the most important evidence for the establishment of General Relativity (GR) \cite{Dyson:1920cwa}. Based on this, the gravitational lensing (GL) phenomenon has been developed into a powerful tool in astronomy. GL can not only link the properties of the source, the lens and the messengers to the observables \cite{Sharon:2014ija,Peng:2006ew,Bartelmann:1999yn,Ade:2015zua,Refregier:2003ct,Lewis:2006fu,Metcalf:2001ap, Hoekstra:2008db}, but also be used to test gravitational theories beyond GR \cite{Keeton:2005jd,Joyce:2016vqv}.

With the discovery of supernova neutrinos \cite{Hirata:1987hu, Bionta:1987qt} and blazer neutrinos  \cite{IceCube:2018dnn,IceCube:2018cha}, the gravitational wave \cite{Abbott:2016blz,Abbott:2016nmj,Abbott:2017oio,TheLIGOScientific:2017qsa, Monitor:2017mdv}, and the more historical cosmic rays \cite{LetessierSelvon:2011dy}, people become more and more interested in the deflection and GL of massive signals. Among these, cosmic rays with ultra-high energies are known to be composed of charged massive particles of protons and  heavier nuclei \cite{AlvesBatista:2019tlv}. Therefore, not only are they massive but carry charge and experience electric interaction if the spacetime is also charged. 

Previously, there have been a large amount of studies on the trajectory of charged signals in electrically or magnetically charged spacetimes, including the Reissner-Nordstr\"{o}m (RN) spacetime in the former case
\cite{SAKA, Pugliese:2011py, Das:2016opi,Pugliese:2013xfa}, and some particular spacetime background supplemented by a weak magnetic field in the latter \cite{Abdujabbarov:2009az,Zahrani:2013up,Tursunov:2016dss}. However, most of these are either concerned with the circular motion of the signal or the general features of the motion.  Concentrating on the deflection and GL of charged signals, then they have been studied in the weak field limit in arbitrary static and spherically symmetric (SSS) spacetimes using perturbative method  \cite{Xu:2021rld} and in particular spherically or axially symmetric spacetimes using Gauss-Bonnet theorem method \cite{Crisnejo:2019xtp,Jusufi:2019rcw,Li:2020ozr,Li:2021xhy}. It was shown that in general, the repulsion (or attraction) between small signal and lens charges will increase (or decrease) the deflection angle at the leading order, i.e., at the order $\calco(M/b)$ where $M$ and $b$ are the mass of the lens and impact parameter of the signal respectively. For a gravitational-electric dual lensing in the weak field limit, this implies that in the repulsion (or attraction) case the impact parameters, the apparent angles, the magnifications and total travel times of the images will be smaller (or larger) \cite{Xu:2021rld}. 

However, these results are more or less intuitively expected because in the weak field limit, gravity resembles its Newtonian limit and therefore both the gravitational and electric potentials take the inverse square form. Consequently the physics due to both interactions can be expected in the same way, except a possible choice of signs of the electric interaction. In contrast, in the strong field limit (SFL), GR deviates most from its Newtonian limit but the Coulomb potential keeps its form. Then one would naturally anticipate more interesting interplay or competition between these two kinds of interactions and more nontrivial results are expected. 

The signal deflection and GL in the SFL have been studied for a relatively long time. 
Virbhadra et al. first pointed out the existence of the relativistic image sequences \cite{Virbhadra:1999nm, Virbhadra:2002ju}.
Bozza et al. systematically studied the GL in the SFL for null signals in SSS spacetimes and in equatorial plane of stationary and axially symmetric spacetimes \cite{Bozza:2002zj,Bozza:2007gt,Bozza:2009yw}, and then  many authors followed in particular spacetimes  \cite{Perlick:2004tq,Whisker:2004gq, Bozza:2005tg,Eiroa:2005ag,Nandi:2006ds, Chen:2009eu,Stefanov:2010xz,Tsukamoto:2012xs, Sahu:2012er, Wei:2014dka,Tsukamoto:2016oca}. In Ref. \cite{Jia:2020qzt,Liu:2021ckg} we were able to develop a perturbative method applicable to arbitrary SSS spacetime to study the deflection, GL and time delay for neutral timelike signals. It was found that in the SFL, i.e. as the impact parameter $b$ approaches its critical value $b_c$, the deflection takes a simple quasi-power series form
\be
\Delta\phi=\sum_{n=0}\lsb -C_n\ln \lb 1-\frac{b_c}{b}\rb+D_n\rsb  \lb 1-\frac{b_c}{b}\rb^n,
\ee
and the time delay has a simple physical interpretation. In this work, we would like to examine how the electric interaction will affect these results, both qualitatively and quantitatively. We will show that our method not only works for charged signals in arbitrary SSS charged spacetime, but it can automatically take into account the finite distance effect of the source and observer. 

The paper is organized as follows. In Sec. \ref{sec:eomps} we lay out the general equations involved in the problem. In Sec. \ref{sec:pesfl}, the perturbative method to compute the deflection angle and total travel time with electric interaction is presented. In Sec. \ref{sec:dphirn}, we take the RN spacetime as an example to show how the formalism works in a particular spacetime. The corresponding GL and black hole (BH) shadow are analyzed in Sec. \ref{sec:glsflcs}, where the images' apparent angles, their magnifications, time delays as well as the BH shadow sizes are solved. 
Sec. \ref{sec:disc} closed the work by a discussion. Throughout the paper we work in the units where  $G=c=1/(4\pi\epsilon_0)=1$.

\section{Metric, motion equation and particle sphere with electric interaction \label{sec:eomps}}

The most general SSS metric can be described by the line element
\begin{equation}\label{1}
	\mathrm{d}s^{2}=-A(r)\mathrm{d}t^{2}+B(r)\mathrm{d}r^{2}+C(r)(\mathrm{d}\theta^{2}+\sin ^{2}\theta \mathrm{d}\phi^{2}),
\end{equation}
where $t,~r,~\theta,~\phi$ are coordinates and $A(r),~B(r)$ and $C(r)$ are functions of the radial coordinate only. 

When the signal is charged and the electromagnetic force is taken into account, the signal will not follow the geodesic but the Lorentz equation \cite{Rohrlich}
\begin{equation}\label{2}
	\frac{\mathrm{d}^{2}x^{\rho}}{\mathrm{d}\tau^{2}}+\Gamma^{\rho}_{\mu\nu}\frac{\mathrm{d}x^{\mu}}{\mathrm{d}\tau}\frac{\mathrm{d}x^{\nu}}{\mathrm{d}\tau}=\frac{q}{m}F^{\rho}_{\mu}\frac{\mathrm{d}x^{\mu}}{\mathrm{d}\tau},
\end{equation}
where $q$ and $m$ are the charge and mass of the signal and $\tau$ is the proper time. The electromagnetic field strength $F_{\mu\nu}$ is given by 
\be F_{\mu\nu}=\partial_{\mu}A_{\nu}-\partial_{\nu}A_{\mu},\ee
where $A_\mu$ is the four potential of the electromagnetic field. For simplicity, in this work we will only consider the effect of an electric field but not the magnetic field. That is, we assume $A_0(r)\neq 0$ and $A_i=0~(i=1,2,3)$.
Without losing any generality, for such a SSS spacetime and electric potential we can always assume that the trajectory lies in the equatorial plane so that $\theta(\tau)=\pi/2$, and then Eq. \eqref{2} after the first integrals reduces to three separate equations 
\begin{subequations}\label{eq:eoms}
	\begin{align}
		& \dot{t}=\frac{E+qA_{0}}{mA},\label{3a}                \\
		& \dot{\phi}=\frac{L}{mC},\label{3b}                    \\
		& \dot{r}^{2}=\frac{\left[(E+qA_{0})^{2}-m^{2}A\right]C-L^{2}A}{m^{2}ABC},\label{3c}
	\end{align}
\end{subequations}
where $E$ and $L$ are integration constants that can be interpreted as the energy and the angular momentum of the particle respectively. Comparing to the case without the electric interaction (setting $q$ to zero), we observe that the changes in these equations happen in and only in terms containing $E$: all $E$ is replaced by the $E+qA_0(r)$, i.e., the electric potential energy manifests itself in this way.

In asymptotically flat spacetimes, on which this work will concentrate, $L$ and $E$ can be related to the impact parameter $b$ and asymptotic velocity $v$ of the signal 
\begin{subequations}\label{eq:ledefs}
	\begin{align}
		& L=\left| \mathbf{r} \times \mathbf{p} \right|=\frac{mv}{\sqrt{1-v^{2}}}b,\label{5a} \\
		& E=\frac{m}{\sqrt{1-v^{2}}}.\label{5b}
	\end{align}
\end{subequations}
Moreover, $L$ can also be related to the shortest approach $r_0$ of the trajectory, which is defined through $\dot{r}|_{r=r_0}=0$. Using Eq. \eqref{3c}, the equation determining $r_0$ becomes
\begin{equation}\label{4}
	L=\sqrt{[(E+qA_{0}(r_0))^{2}-m^{2}A(r_0)]C(r_0)/A(r_0)}. 
\end{equation}
Eqs. \eqref{5a} and \eqref{4} establish the following correspondence between  $b$ and $r_0$,
\begin{equation}\label{7}
	\frac{1}{b}=\frac{\sqrt{E^{2}-m^{2}}}{\sqrt{(E+qA_{0}(r_0))^{2}-m^{2}A(r_0)}}\sqrt{\frac{A(r_0)}{C(r_0)}},
\end{equation}
and this allows us in principle to express the deflection angle and time delay in either of them. 

\begin{figure}
	\includegraphics[width=0.45\textwidth]{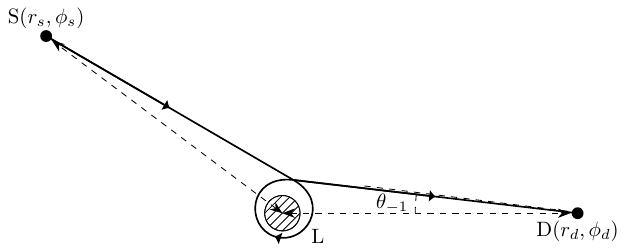}
	\caption{The trajectory deflection and GL in the SFL. The source and detector are located at coordinates $(r_s,\phi_s)$ and $(r_d,\phi_d)$ respectively. The apparent angle in this example is $\theta_{-1}$. \label{fig:glillus}}
\end{figure}

For a signal from a source located at radius $r_s$ to an observer at $r_d$ (see Fig. \ref{fig:glillus}), from Eqs. \eqref{eq:eoms} we see then the change of the angular coordinate $\Delta\phi$ and the total travel time $\Delta t$ become respectively
\begin{subequations}\label{eq:dphidtdefori}
	\begin{align}
		\Delta\phi=&\left[\int_{r_0}^{r_s}+\int_{r_0}^{r_d}\right]\frac{\dd\phi}{\dd r}~\dd r\nn\\
		=&\left[\int_{r_0}^{r_s}+\int_{r_0}^{r_d}\right]\frac{L\sqrt{B/C} }{\sqrt{[(E+qA_{0})^2/A-m^2]C-L^{2}}}\mathrm{d}r, \label{8}\\
		\Delta t=&\left[\int_{r_0}^{r_s}+\int_{r_0}^{r_d}\right]\frac{\dd t}{\dd r}~\dd r\nn\\
		=&\left[\int_{r_0}^{r_s}+\int_{r_0}^{r_d}\right]\frac{(E+qA_0)\sqrt{BC}}{LA} \nonumber\\
		&\times\frac{L }{\sqrt{[(E+qA_{0})^2/A-m^2]C-L^{2}}}\mathrm{d}r. \label{8p5}
	\end{align}
\end{subequations}
For general SSS spacetime metric and potential $A_0(r)$, these integrals usually can not be worked out analytically even for the simplest RN case. Therefore, some special techniques are required to approximate them. The technique we will use in this work is actually a generalization of the perturbative method developed in Ref. \cite{Jia:2020qzt,Liu:2021ckg} for neutral signals to the current case, in which the electric interaction is now taken into account. 

In the SFL, the spacetime allows the existence of some critical value $r_{0c}$ of the closest approach $r_0$. When $r_0\to r_{0c}^+$, the trajectory will circulate many times around the center before propagating to the observer; while below it, the signal will enter the event horizon and therefore not reach the observer anymore. 
Previously, the existence of such $r_{0c}$ has been known for many BH spacetimes for photons (the photon sphere) \cite{Bozza:2001xd}, and for massive signals too (the particle sphere (PS)) \cite{Jia:2020qzt,Wang:2019rvq}. In this work, because of the presence of the electric interaction, the existence and size of $r_{0c}$ might both be affected. Therefore, next, we derive the equation determining $r_{0c}$ in this new case first. This process will also help us to better understand its dependence on various parameters of the spacetime and signal.

We first re-write Eq. \eqref{3c} into the following form 
\begin{align}
	&\frac{1}{2}\lb m\frac{\dd r}{\dd\tau}\rb ^{2}\nn\\
	&+\frac{\lsb L^{2}A+\lb m^2A-q^2A_0^2-2qEA_0\rb C\rsb}{2ABC}=\frac{E^2}{2AB} .\label{eq:effp}
\end{align}
Then we can define the second term on the left hand side as the effective potential
\be 
V_{\mathrm{eff}}=\frac{L^{2}A+\lb m^2A-q^2A_0^2-2qEA_0\rb C}{2ABC} .
\label{eq:effpdef}\ee
When $A=1/B$ and $A_0=0$, as in many well-known SSS chargeless spacetimes, this reduces to the effective potential in the corresponding spacetimes \cite{Wang:2019rvq}. The terms proportional to $q^2A_0^2$ and $qEA_0$ in Eq. \eqref{eq:effpdef} are due to the electric interaction. From Eq. \eqref{eq:effp} we see that the allowed range of $r$ for a signal is when $V_{\mathrm{eff}}(r)$ is smaller than the right hand side of Eq. \eqref{eq:effp}, i.e.,
\be 
V_{\mathrm{eff}}(r)\leq \frac{E^2}{2A(r)B(r)}. \label{eq:veffeqe}
\ee
The equal sign of this equation is actually equivalent to Eq. \eqref{4} and therefore also determines the closest radius $r_0$.

\begin{figure}[htp!]
	\includegraphics[width=0.45\textwidth]{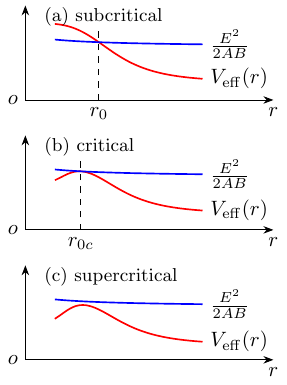}
	\caption{The effective potential $V_{\mathrm{eff}}(r)$ and $E^2/(2AB)$ as functions of $r$ for different values of parameters in them. (a) The case with a regular trajectory bouncing back at $r_0$. (b) The case that the trajectory experiencing the critical $r_{0c}$. (c) The supercritical trajectory case, for which the $r_0$ does not exist anymore. \label{fig:effpillus}}
\end{figure}

Now for many interesting spacetimes, their Eq. \eqref{eq:effp} also possesses a critical behavior. As certain parameter(s) in $V_{\mathrm{eff}}(r)$ and $E^2/(2AB)$ varies (e.g., $L$ or equivalently $b$, $q$ and $Q$, etc), if these two quantities become barely intersecting and eventually detach from each other (see Fig. \ref{fig:effpillus}), then at this point, the $r_0$, which is defined by the equal sign of Eq. \eqref{eq:veffeqe}, ceases to exist and the signal coming from larger $r$ will be able to continue to smaller $r$. This critical behavior can be thought to happen at the critical value of the above mentioned parameters. But on the other hand, since this detaching point in the radial direction is also the location of $r_0$, it can also be thought as the critical value of $r_0$, which we will denote as $r_{0c}$. Mathematically, $r_{0c}$ can be determined from the conditions 
\begin{subequations}
	\label{eq:r0cfixcond}
	\begin{align}
		&\left.\frac{\dd \lsb V_{\mathrm{eff}}(r)-E^2/(2A(r)B(r))\rsb}{\dd r}\right|_{r=r_{0c}}=0,\label{eq:detach}\\
		&V_{\mathrm{eff}}(r_{0c})= \frac{E^2}{2A(r_{0c})B(r_{0c})}.\label{eq:r0cdef}
	\end{align}
\end{subequations}
And we can also supplement the condition
\be 
V_{\mathrm{eff}}(r)< \frac{E^2}{2A(r)B(r)}~~\text{for}~~
r>r_{0c}
\ee 
to ensure the signal can come from large $r$. 
Once the spacetime metric is specified, Eq. \eqref{eq:r0cfixcond} will fix the value of $r_{0c}$ completely. Using Eq. \eqref{7}, the corresponding critical impact parameter $b_c$ becomes
\begin{equation}\label{13}
	b_c=\frac{\sqrt{(E+qA_{0}(r_{0c}))^{2}-m^{2}A(r_{0c})}}{\sqrt{E^{2}-m^{2}}}\sqrt{\frac{C(r_{0c})}{A(r_{0c})}}.
\end{equation}

\section{Perturbative expansion in the SFL\label{sec:pesfl}}

To compute the deflection angle and total travel time defined in Eq. \eqref{eq:dphidtdefori} in the SFL, in the following we will extend a perturbative method developed previously in Ref. \cite{Jia:2020qzt,Liu:2021ckg} to the case with electric interaction. It turns out that the methodology and general form of $\Delta\phi$ and $\Delta t$ will also work in this case after some modification of the critical $r_{0c}$ and the expansion coefficients in Eq. \eqref{eq:yfirstfew}. 

In this method, we first define a function $p(x)$ inspired by Eq. \eqref{13}
\begin{equation}\label{14}
	p\lb x\rb=\frac{1}{b_c}-\frac{\sqrt{E^{2}-m^{2}}}{\sqrt{(E+qA_{0}(1/x))^{2}-m^{2}A(1/x)}}\sqrt{\frac{A(1/x)}{C(1/x)}}.
\end{equation}
Using Eq. \eqref{7}, it is clear then 
\be p\lb \frac{1}{r_0}\rb=\frac{1}{b_c}-\frac{1}{b}.\label{eq:poor0}\ee
Denoting the inverse function of $p(x)$ as $w(x)$, the above means
\begin{equation}\label{15}
	\frac{1}{r_0}=w\left(\frac{1}{b_c}-\frac{1}{b}\right)=w\left(\frac{1-b_c/b}{b_c}\right).
\end{equation}
Using the function $w(x)$, we can define a change of variables in the integrals \eqref{eq:dphidtdefori} from $r$ to $\xi$, which are connected by
\begin{equation}\label{16}
	\frac{1}{r}=w\left(\frac{\xi}{b_c}\right)\ ,\ \mathrm{i.e.},\ p\left(\frac{1}{r}\right)=\frac{\xi}{b_c}.
\end{equation}
With this, the integral limits and integrands of Eq. \eqref{eq:dphidtdefori} change in the following way
\begin{subequations}\label{eq:17all}
	\begin{align}
		& r_0\rightarrow 1-b_c/b,\label{17a}               \\
		& r_{s,d}\rightarrow 1-b_c\frac{\sqrt{E^{2}-m^{2}}}{\sqrt{(E+qA_{0}(r_{s,d}))^{2}-m^{2}A(r_{s,d})}}\sqrt{\frac{A(r_{s,d})}{C(r_{s,d})}}   \nn \\
		&~~~~~~~~\equiv \eta_{s,d},\label{17b} \\
		& \mathrm{d}r\rightarrow -\frac{w'}{b_cw^{2}}\mathrm{d}\xi,\label{17c}  \\
		& A(r)\rightarrow A(1/w),\ B(r)\rightarrow B(1/w),\ C(r)\rightarrow C(1/w),\notag\\
		&A_{0}(r)\rightarrow A_{0}(1/w),\label{17d}     \\
		& \frac{L}{\sqrt{[(E+qA_{0}(r))^{2}/A(r)-m^{2}]C(r)-L^{2}}}   \nn    \\
		& \rightarrow \frac{\xi-1}{\sqrt{2-a-\xi}\sqrt{\xi-a}},\label{17e}
	\end{align}
\end{subequations}
where 
\be 
w=w\left(\frac{\xi}{b_c}\right)\  \text{and} \ w'=\left.\frac{\mathrm{d}[w(x)]}{\mathrm{d}x} \right|_{x=\xi/b_c},\ee 
and in Eq. \eqref{17b} we have defined $\eta_{s,d}$, and in Eq. \eqref{17e} and henceforth we set 
\be 
a=1-b_c/b. \label{eq:adef} \ee
Collecting terms in Eq. \eqref{eq:17all} together and using Eq. \eqref{eq:ledefs} and \eqref{eq:adef}, Eq. \eqref{eq:dphidtdefori} becomes
\begin{subequations}
	\label{eq:dphidtdef}
	\begin{align}\label{18}
		\Delta\phi=&\sum_{i=s,d}\int_{a}^{\eta_{i}}
		\sqrt{\frac{B(1/w)}{C(1/w)}}\frac{w'(\xi-1)}{w^{2}b_c} \frac{1}{\sqrt{2-a-\xi}}\frac{1}{\sqrt{\xi-a}}\mathrm{d}\xi.\\
		\Delta t=&\sum_{i=s,d}\int_{a}^{\eta_{i}}
		\frac{[1+qA_{0}(1/w)/E]\sqrt{B(1/w)C(1/w)}}{A(1/w)}\notag\\
		&~~~\times\frac{w'(1-a)(\xi-1)}{vw^{2}b_{c}^{2}}\frac{1}{\sqrt{2-a-\xi}} \frac1{\sqrt{\xi-a}}\mathrm{d}\xi.
		\label{18p5}
	\end{align}
\end{subequations}

In the SFL, i.e., $r_0\rightarrow r_{0c}^{+}$, $b\rightarrow b_c^{+}$ and $a\rightarrow 0^{+}$, the integrands of Eq. \eqref{eq:dphidtdef} can be perturbatively expanded for small $\xi$ and therefore it allows us to find an effective approximation of $\Delta\phi$ and $\Delta t$. 
Among the four factors of each of the integrands, the last one is the simplest and can be kept since its product with a power series of $\xi$ can be directly integrated. The third factor has a simple Taylor expansion
\begin{equation}\label{20}
	\frac{1}{\sqrt{2-a-\xi}}=\sum_{n=0}^{\infty}\frac{(2n-1)!!}{(2n)!!}\frac{\xi^{n}}{(2-a)^{n+\frac{1}{2}}}.
\end{equation}
Denoting the product of the first two factors of Eq. \eqref{18} as $f(\xi)$ and those of Eq. \eqref{18p5} as $g(\xi)$, we have
\begin{align}\label{21}
	f(\xi)=&\sqrt{\frac{B(1/w)}{C(1/w)}}\frac{w'(\xi-1)}{w^{2}b_c},\\
	g(\xi)=&\frac{[1+qA_{0}(1/w)/E]\sqrt{B(1/w)C(1/w)}}{A(1/w)}\nn\\
	&\times\frac{w'(1-a)(\xi-1)}{vw^{2}b_{c}^{2}},\label{21p5}
\end{align} 
we then are able to show that $f(\xi)$ and $g(\xi)$ have the following form of expansions
\begin{subequations}\label{eq:fgexp}
	\begin{align}\label{22}
		f(\xi)&=\sum_{n=-1}^{\infty} f_{n}\xi^{\frac{n}{2}},\\
		g(\xi)&=(1-a)\sum_{n=-1}^{\infty} g_{n}\xi^{\frac{n}{2}}.
		\label{22p5}
	\end{align}
\end{subequations}
Note in these expansions, the series expansion of $w(x)$ is needed, and it can always be obtained using the Lagrange Inversion Theorem from the series form of its inverse function $p(x)$. In Eq. \eqref{eq:fgexp}, the initial summation index is -1 and the powers are half integers because $q'(x)$ has a singularity at $x=0$. 
We can indeed collectively denote the product of the first three factors in Eq. \eqref{18} as $y(\xi,a)$ and that of Eq. \eqref{18p5} as $z(\xi,a)$
\begin{subequations}\label{eq:yzdefsini}
	\begin{align}\label{19}
		y(\xi,a)=&\sqrt{\frac{B(1/w)}{C(1/w)}}\frac{w'(\xi-1)}{w^{2}b_c} \frac{1}{\sqrt{2-a-\xi}},\\
		z(\xi,a)=&\frac{[1+qA_{0}(1/w)/E]\sqrt{B(1/w)C(1/w)}}{A(1/w)}\nn\\
		&\times\frac{w'(1-a)(\xi-1)}{vw^{2}b_{c}^{2}} \frac{1}{\sqrt{2-a-\xi}}. \label{19p5}
	\end{align}
\end{subequations}
Then their expansions are given by the product of expansions \eqref{20} and \eqref{22}, and \eqref{20} and \eqref{22p5}. After some re-organization, they become respectively  
\begin{subequations}\label{eq:yzexp}
	\begin{align}\label{23}
		y(\xi)&=\sum_{n=-1}^{\infty}\left[ \sum_{m=0}^{\left[ \frac{n+1}{2} \right] } \frac{a^{m}}{(2-a)^{\left[ \frac{n+1}{2}\right]+\frac{1}{2}}}y_{n,m}\right] \xi^{n/2},\\
		z(\xi)&=\sum_{n=-1}^{\infty}\left[ \sum_{m=0}^{\left[ \frac{n+1}{2} \right] } \frac{(1-a)a^{m}}{(2-a)^{\left[ \frac{n+1}{2}\right]+\frac{1}{2}}}z_{n,m}\right] \xi^{n/2},\label{23p5}
	\end{align}
\end{subequations}
where $y_{n,m}$ and $z_{n,m}$ denote coefficients of corresponding powers of $\xi$ and $a$. Note that these $y_{n,m}$ and $z_{n,m}$ can be completely determined once the metric functions are fixed. The first few of them for general SSS spacetime with a critical $r_{0c}$ are given in Eqs. \eqref{eq:yfirstfew} and \eqref{eq:zfirstfew}. 

Substituting Eq. \eqref{eq:yzexp} into Eq. \eqref{eq:dphidtdef}, $\Delta\phi$ and $\Delta t$ are written as series of integrals of $\xi$
\begin{subequations}
	\label{eq:dphidtseries}
	\begin{align}\label{24}
		\Delta\phi&=\sum_{i=s,d} \sum_{n=-1}^{\infty} \sum_{m=0}^{\left[\frac{n+1}{2}\right]}\frac{a^m}{(2-a)^{\left[\frac{n+1}{2}\right]+\frac{1}{2}}} y_{n,m} \int_{a}^{\eta_{i}} \frac{\xi^{n/2}}{\sqrt{\xi-a}} \mathrm{d}\xi,\\
		\Delta t&=\sum_{i=s,d} \sum_{n=-1}^{\infty} \sum_{m=0}^{\left[\frac{n+1}{2}\right]}\frac{(1-a)a^m}{(2-a)^{\left[\frac{n+1}{2}\right]+\frac{1}{2}}} z_{n,m} \int_{a}^{\eta_{i}} \frac{\xi^{n/2}}{\sqrt{\xi-a}} \mathrm{d}\xi.
	\end{align}
\end{subequations}
We emphasize that the integrals in the above series can always be carried out and the results are some elementary functions presented in Eqs. (A1) of Ref. \cite{Jia:2020qzt}.  
Using these formulas, $\Delta\phi$ and $\Delta t$ are finally found to be 
\begin{subequations}
	\label{eq:dphidtf1}
	\begin{align}
		\Delta\phi= & \sum_{i=s,d} \sum_{k=0}^{\infty} \sum_{m=0}^{k} \frac{a^{m}}{(2-a)^{k+\frac{1}{2}}} \Bigg\{ y_{2k-1,m}\cdot \frac{a^{k}C^{k}_{2k}}{4^{k}}\Bigg[-\mathrm{ln}a  \nn   \\
		& +2\mathrm{ln} \left(\sqrt{\eta_{i}}+\sqrt{\eta_{i}-a}\right)+\sum^{k}_{j=1}\frac{4^{j}}{jC^{j}_{2j}}\lb \frac{\eta_{i}}{a}\rb^{j}\sqrt{1-\frac{a}{\eta_{i}}}\Bigg]   \nn \\
		& +y_{2k,m}\sum^{k}_{j=0}\frac{2C^{k}_{j}a^{k-j}(\eta_{i}-a)^{j+1/2}}{2j+1} \Bigg\},     \label{25}\\
		\Delta t= & \sum_{i=s,d} \sum_{k=0}^{\infty} \sum_{m=0}^{k} \frac{(1-a)a^{m}}{(2-a)^{k+\frac{1}{2}}} \Bigg\{ z_{2k-1,m}\cdot \frac{a^{k}C^{k}_{2k}}{4^{k}}\Bigg[-\mathrm{ln}a  \nn   \\
		& +2\mathrm{ln} \left(\sqrt{\eta_{i}}+\sqrt{\eta_{i}-a}\right)+\sum^{k}_{j=1}\frac{4^{j}}{jC^{j}_{2j}}\lb \frac{\eta_{i}}{a}\rb^{j}\sqrt{1-\frac{a}{\eta_{i}}}\Bigg]   \nn \\
		& +z_{2k,m}\sum^{k}_{j=0}\frac{2C^{k}_{j}a^{k-j}(\eta_{i}-a)^{j+1/2}}{2j+1} \Bigg\}.     \label{25p5}
	\end{align}
\end{subequations}

A few remarks are in order here. First of all, for both $\Delta\phi$ and $\Delta t$, we observe that when $r_s$ and $r_d$ are not infinite, there is only one divergence  proportional to $\ln a$ in the SFL $a\to 0^+$, which is only contained by the $k=m=0$ term. Secondly, we can further expand other functions involving $a$, i.e. $\ln(\sqrt{\eta_i}+\sqrt{\eta_i-a})$ and $\sqrt{1-a/\eta_i}$ etc., in the limit $a\to 0^+$, and the results should become quasi-power series of $a$ with the coefficient  of $a^n$ containing one $\ln a$. That is,
\begin{subequations}\label{eq:dphidtform}
	\begin{align}\label{26}
		\Delta\phi&=\sum_{n=0}^{\infty}\left[ -C_{n}\ln a+D_{n}\right]a^{n},\\
		\Delta t&=\sum_{n=0}^{\infty}\left[ -C_{n}^\prime\ln a+D_{n}^\prime\right]a^{n},
	\end{align}
\end{subequations}
where $C_{n}$ and $D_{n}$, and $C_{n}^\prime$ and $D_{n}^\prime$ are constants determined by the initial conditions of the trajectory and metric function parameters. Thirdly, from Eq.\eqref{25}, it is seen that to the $\mathcal{O} (a)^{0}$ order, the $m=0,~j=k$ terms contribute dominantly to $\Delta\phi$ so that it becomes
\bea
\Delta\phi & =& \sum_{i=s,d}\left\{ \frac{\sqrt{2}y_{-1,0}}{2}\left[ -\ln a+2\ln (2\sqrt{\eta_{i}})\right]\right.   \nn                 \\
&& \left.+\sum_{n=0}^{\infty}\frac{2y_{n,0}\eta_{i}^{\frac{n+1}{2}}}{2^{\left[\frac{n+1}{2}\right]+\frac{1}{2}}(n+1)}\right\}+\mathcal{O}(a)^{1} \nn \\
& \equiv & -C_{0}\ln a+D_{0}(\eta_{s},\eta_{d})+\mathcal{O}(a)^{1}, \label{27b}
\eea
where we have identified
\bea\label{28}
C_{0}&=&\sqrt{2}y_{-1,0},\\
\label{29}
D_{0}&=&\sum_{i=s,d}\left[ \sqrt{2}y_{-1,0}\ln (2\sqrt{\eta_{i}})+\sum_{n=0}^{\infty}\frac{2y_{n,0}\eta_{i}^{\frac{n+1}{2}}}{2^{\left[\frac{n+1}{2}\right]+\frac{1}{2}}(n+1)}\right]. \notag\\
&
\eea
The travel time \eqref{25p5} can also be similarly expanded again for small $a$ so that to the leading orders it becomes
\bea
\Delta t & =& \sum_{i=s,d}\left\{ \frac{\sqrt{2}z_{-1,0}}{2}\left[ -\ln a+2\ln (2\sqrt{\eta_{i}})\right]\right.   \nn                 \\
&& \left.+\sum_{n=0}^{\infty}\frac{2z_{n,0}\eta_{i}^{\frac{n+1}{2}}}{2^{\left[\frac{n+1}{2}\right]+\frac{1}{2}}(n+1)}\right\}+\mathcal{O}(a)^{1} \nn \\
& \equiv & -C'_{0}\ln a+D'_{0}(\eta_{s},\eta_{d})+\mathcal{O}(a)^{1} , \label{27bp5}
\eea
where we see that
\bea
C'_{0}&=&\sqrt{2}z_{-1,0},\label{C0'}\\
D'_{0}&=&\sum_{i=s,d}\left[ \sqrt{2}z_{-1,0}\ln (2\sqrt{\eta_{i}})+\sum_{n=0}^{\infty}\frac{2z_{n,0}\eta_{i}^{\frac{n+1}{2}}}{2^{\left[\frac{n+1}{2}\right]+\frac{1}{2}}(n+1)}\right].\notag\\
&\label{27bp6}
\eea

At this point, it is appropriate to point out the difference between the perturbative method and result in this work and the case without electric interaction \cite{Jia:2020qzt,Liu:2021ckg}. Superficially, the steps from Eq. \eqref{eq:poor0} to the result \eqref{27bp6} are very similar to the case without electric interaction. However, indeed there exist a fundamental difference in the definition of $p(x)$  in Eq. \eqref{13}, where the electric potential explicitly contributes in the current case but not so in Ref. \cite{Jia:2020qzt,Liu:2021ckg}. Consequently, its inverse function $w(x)$ is also changed by $A_0(r)$ and this further changed all formulas involving the expansions of terms containing $w(\xi/b_c)$. More specifically, the coefficients $f_n$ and $g_n$ in Eq. \eqref{eq:fgexp} and consequently $y_{n,m}$ and $z_{n,m}$ in Eqs. \eqref{eq:yzexp}-\eqref{eq:dphidtf1}, as well as the $C_n,~D_n,~C_n^\prime,~D_n^\prime$ in Eqs. \eqref{eq:dphidtform}-\eqref{27bp6}, all contain contribution from the electric potential. We now show their dependence on $A_0(r)$, as well as other metric functions, more explicitly. 

Assuming the metric functions and the electric potential have the following series expansions near $r_{0c}$
\begin{subequations}\label{eq:metrica0exp}
	\begin{align}
		&A(r\rightarrow r_{0c})=\sum_{n=0}^{\infty}a_{n}(r-r_{0c})^{n},\label{30a} \\
		&B(r\rightarrow r_{0c})=\sum_{n=0}^{\infty}b_{n}(r-r_{0c})^{n},\label{30b} \\
		&C(r\rightarrow r_{0c})=\sum_{n=0}^{\infty}c_{n}(r-r_{0c})^{n},\label{30c} \\
		&A_{0}(r\rightarrow r_{0c})=\sum_{n=0}^{\infty}a_{0n}(r-r_{0c})^{n}\label{30d},
	\end{align}
\end{subequations}
where $a_n,~b_n,c_n$ and $a_{0n}~(n=0,~1,~\cdots)$ are the coefficients,
and substituting them into Eqs. \eqref{14}, we can solve the series $p(x)$. Inverting it to find $w(x)$ and then further substituting into Eq. \eqref{eq:yzdefsini}, we can compute $y_{n,m}$ and $z_{n,m}$ to any desired order. The first two of $y_{n,m}$, i.e. $y_{-1,0}$ in Eq. \eqref{28} which fixes the divergence of $\Delta\phi$ and $y_{0,0}$ in Eq. \eqref{29} which dominates the constant term of $\Delta\phi$, are 
\begin{subequations}\label{eq:yfirstfew}
	\begin{align}
		& y_{-1,0}=b_c\sqrt{\frac{b_{0}}{2c_{0}T_{2}}},\label{32a}      \\
		& y_{0,0}=\frac{b_c^{2}(b_{1}c_{0}T_{2}-b_{0}c_{1}T_{2}-2b_{0}c_{0}T_{3})}{2\sqrt{b_{0}}c_{0}^{3/2}T_{2}^{2}},\label{32b}
	\end{align}
\end{subequations}
where $b_c$ is given in Eq. \eqref{13} and $T_{2}$ and $T_{3}$ are related to the first few coefficients in Eq. \eqref{eq:metrica0exp} by
\begin{widetext}
	\begin{align}
		T_{2}= & \frac{1}{E^2-m^2} \bigg\{c_0 \left[\frac{2 a_{02} q \left(a_{00}
			q+E\right)+a_{01}^2 q^2}{a_0}-\frac{ 2a_1 a_{01} q \left(a_{00} q+E\right)+a_2 \left(a_{00}
			q+E\right){}^2}{a_0^2}+\frac{a_1^2 \left(a_{00} q+E\right){}^2}{a_0^3}\right]\nn\\
		&\left.+c_1 \left[\frac{2 a_{01} q\left(a_{00} q+E\right)}{a_0}-\frac{a_1 \left(a_{00} q+E\right){}^2}{a_0^2}\right]+\frac{c_2 \left[\left(a_{00} q+E\right){}^2-a_0 m^2\right]}{a_0}\right\}              ,  \label{33}     \\           
		T_{3}= & \frac{1}{E^2-m^2}  \bigg\{ c_0 \left[\frac{2 q \left(a_{03}
			\left(a_{00} q+E\right)+a_{01} a_{02} q\right)}{a_0}\right.\nn\\
		&-\frac{a_3 \left(a_{00}
			q+E\right){}^2+2 a_2 a_{01} q \left(a_{00} q+E\right)+a_1 q \left(2 a_{02} \left(a_{00} q+E\right)+a_{01}^2 q\right)}{a_0^2}\nn\\
		&\left.+\frac{2 a_1 \left(a_{00} q+E\right) \left(a_2 \left(a_{00} q+E\right)+a_1 a_{01} q\right)}{a_0^3}-\frac{a_1^3 \left(a_{00} q+E\right){}^2}{a_0^4}\right]\nn\\
		&+c_1 \left[\frac{q \left(2 a_{02} \left(a_{00} q+E\right)+a_{01}^2 q\right)}{a_0}-\frac{a_2 \left(a_{00} q+E\right){}^2+2 a_1 a_{01} q \left(a_{00} q+E\right)}{a_0^2}+\frac{a_1^2 \left(a_{00}
			q+E\right){}^2}{a_0^3}\right]        \nn   \\
		&\left. +c_2 \left[\frac{2 a_{01} q \left(a_{00} q+E\right)}{a_0}-\frac{a_1 \left(a_{00} q+E\right){}^2}{a_0^2}   \right]
		+c_3 \left[-m^2+\frac{ \left(a_{00} q+E\right){}^2}{a_0} \right]\right\}.   \label{34}
	\end{align}
	Similarly, the first few $z_{n,m}$ are found to be 
	\begin{subequations}\label{eq:zfirstfew}
		\begin{align}
			z_{-1,0}&=\frac{\left(a_{00} q +E\right)}{\sqrt{2}  E a_0  v }\sqrt{\frac{b_0 c_0}{T_2}},\label{eq:zm10}\\
			z_{0,0}&=\frac{b_c \left[\left(2 q  a_0 a_{01} b_0
				c_0 +   (q a_{00}+E) (a_0 b_1 c_0 +a_0
				b_0 c_1 -2a_1 b_0 c_0) \right)T_2-2 (q a_{00}+E)  a_0  b_0 c_0 T_3\right]}{2  E v a_0^2 \sqrt{b_0 c_0}
				T_2^2}.
		\end{align}
	\end{subequations}
\end{widetext}
The higher order terms of $y_{n,m}$ and $z_{n,m}$  can also be computed but are too long to present here. It is seen from these equations that $a_{0n}$ explicitly appear in all of the $y_{n,m}$ and $z_{n,m}$ and therefore will affect how the deflection angle $\Delta\phi$ and total travel time $\Delta t$ diverge. In Sec. \ref{sec:dphirn}, we will study the dependence of $\Delta\phi$ on the electric interaction using RN spacetime as an example. It is also interesting to note from Eqs. \eqref{33} and \eqref{34} by factoring out an $E^2$ from the curly brackets that both $T_2,~T_3$, and indeed all orders of $y_{n,m}$ and $z_{n,m}$ and consequently the entire $\Delta\phi$ and $\Delta t$, depend on $(q,~m,~E)$ only through the ratios $q/E$ and $v=\sqrt{E^2-m^2}/E$. That is, there is one degree of degeneracy among these three parameters. This fact was also observed in the weak deflection case in Ref. \cite{Xu:2021rld,Li:2021xhy}. 

\section{The RN spacetime case \label{sec:dphirn}}

In this section, We apply our result to the RN spacetime, which is the simplest SSS spacetime that allows both electric and gravitational interactions. 
The metric functions and the electric potential in the RN spacetime are given by
\bea\label{35}
&&A(r)=\frac{1}{B(r)}=1-\frac{2M}{r}+\frac{Q^{2}}{r^{2}},\ C(r)=r^{2}, \nn \\
&&A_{0}(r)=-\frac{Q}{r}. \label{eq:rnmetrica0}
\eea
We will examine the validity of the results in Secs. \ref{sec:eomps} and \ref{sec:pesfl}, particularly concentrating on how the electric interaction will affect the PS radius $r_{0c}$ defined in Eq. \eqref{eq:r0cfixcond}, $b_c$ defined in \eqref{13} and the deflection $\Delta\phi$ in Eq. \eqref{27b}. 

\subsection{The critical $r_{0c}$ and $b_c$}

Substituting Eq. \eqref{eq:rnmetrica0} into  Eq. \eqref{eq:r0cfixcond} and simplifying the result, it is not difficult to find that $r_{0c}$ satisfies the following quartic equation 
\begin{align}
	& \left(E^2-m^2\right)r_{0c}^4 + \lsb (4m^2-3 E^2) M-E q Q\rsb  r_{0c}^3 \nn\\
	&+\lsb -4m^2M^2+4 Eq  M Q+2(E^2 -m^2)Q^2\rsb r_{0c}^2\nn\\
	& +\lsb (4 m^2-q^2) M - 3 E qQ\rsb Q^2 r_{0c}+(q^2-m^2) Q^4 =0. \label{36}
\end{align}
Denoting the coefficients of $r_{0c}^n$ by $g_n$ for $n=0,~\cdots,~4$, this equation can be solved to yield an explicit formula of the only physical root of $r_{0c}$
(see Eq. (A3) of Ref. \cite{Pang:2018jpm} and \cite{qform})
\begin{align}
	r_{0c}=-\frac{g_3}{4 g_4}+S + \frac{1}{2} \sqrt{-2 P_1-\frac{P_2}{S}-4 S^{2}}, \label{eq:r0cgeneral}
\end{align}
where
\begin{align}
	&P_1=\frac{8 g_4 g_2-3 g_3^2}{8 g_4^2}, \nn\\
	&P_2=\frac{8 g_4^{2} g_1-4 g_4 g_3 g_2+g_3^{3}}{8 g_4^{3}}, \nn\\
	&S=\frac{1}{2} \sqrt{-\frac{2 P_1}{3}+\frac{2 \sqrt{\Delta_{0}} \cos \left(\frac{\varphi}{3}\right)}{3 g_4},} \nn\\
	&\varphi=\cos ^{-1}\left(\frac{\Delta_{1}}{2 \sqrt{\Delta_{0}^{3}}}\right), \nn\\
	&\Delta_{0}=12 g_4 g_0-3 g_3 g_1+g_2^{2}, \nn\\
	&\Delta_{1}=-72 g_4 g_2 g_0+27 g_4 g_1^{2}+27 g_3^{2} g_0-9 g_3 g_2 g_1+2 g_2^{3}.\nn
\end{align}

Although Eq. \eqref{eq:r0cgeneral} is not very transparent to see the effects of various parameters such as $M,~Q,~E,~m$ and $q$ on the above $r_{0c}$, it is however not difficult to demonstrate by dividing Eq. \eqref{36} by $m^2M^4$ that $r_{0c}/M$ depends on them only through the ratios $q/m\equiv \hat{q},~Q/M\equiv \hat{Q}$ and $E/m=1/\sqrt{1-v^2}$. In other words, \be r_{0c}=r_{0c}(\hat{q},\hat{Q},v).\label{eq:r0cdep1}\ee 
Moreover, there are a few limits of Eq. \eqref{eq:r0cgeneral} that one can check. The first is its neutral particle limit, which can be obtained by letting $\hat{q}\to 0$ and then the result agrees with Eq. (18) of Ref. \cite{Pang:2018jpm}.
The second is its Schwarzschild limit reached by letting $\hat{Q}\to 0$
\begin{align}
	r_{0c}(\hat{q},\hat{Q}=0,v)=M\left(2+\frac{4}{\sqrt{8v^{2}+1}+1}\right). \label{eq:r0ccq0}
\end{align}
This agrees with Eq. (18) of Ref. \cite{Jia:2015zon}. The third limit is the ultra-relativistic limit of the charged signal, which is simply the limit $E\to \infty$ while holding $m$ finite or equivalently $v\to 1$. The result is 
\begin{align}
	r_{0c}(\hat{q},\hat{Q},v=1)=\frac{3M}{2}\left(1+\sqrt{1-8\hat{Q}^2/9}\right). \label{eq:r0cv1}
\end{align}
Finally, one can also attempt the $v\to 0$ limit of $r_{0c}$, in which case the Eq. \eqref{36} becomes cubic and a root simpler than Eq. \eqref{eq:r0cgeneral} can be obtained (see Eq. (26) of Ref. \cite{Pang:2018jpm} for the neutral case).  

In both the limits \eqref{eq:r0ccq0} and \eqref{eq:r0cv1}, the dependence of $r_{0c}$ on the charge-mass ratio $\hat{q}$ of the charged signal disappears. The disappearance in the former is quite easy to understand, since in a neutral spacetime like the Schwarzschild one, the signal although charged will not experience any electric interaction. While for the second disappearance, one can show by dividing Eq. \eqref{36} by $E^2M^4$ that $r_{0c}$ will depend on $\hat{q}$ only through $\hat{q}/(E/m)=\hat{q}\sqrt{1-v^2}$. That is, Eq. \eqref{eq:r0cdep1} can be further 
transformed into 
\be r_{0c}=r_{0c}(\hat{q}\sqrt{1-v^2},\hat{Q},v). 
\label{eq:r0cdep2}
\ee
Therefore as $v\to 1$, $\hat{q}$ effectively drops out from $r_{0c}$. 

\begin{figure}[htp!]
	\includegraphics[width=0.45\textwidth]{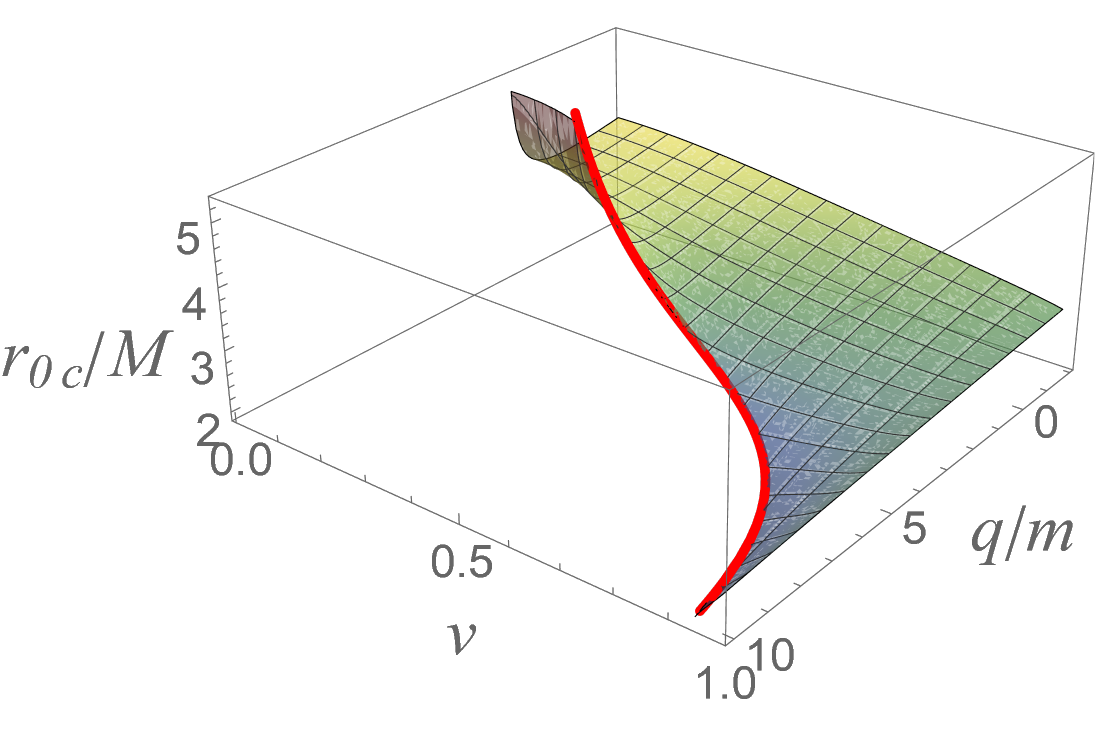}\\
	(a)\\
	\includegraphics[width=0.47\textwidth]{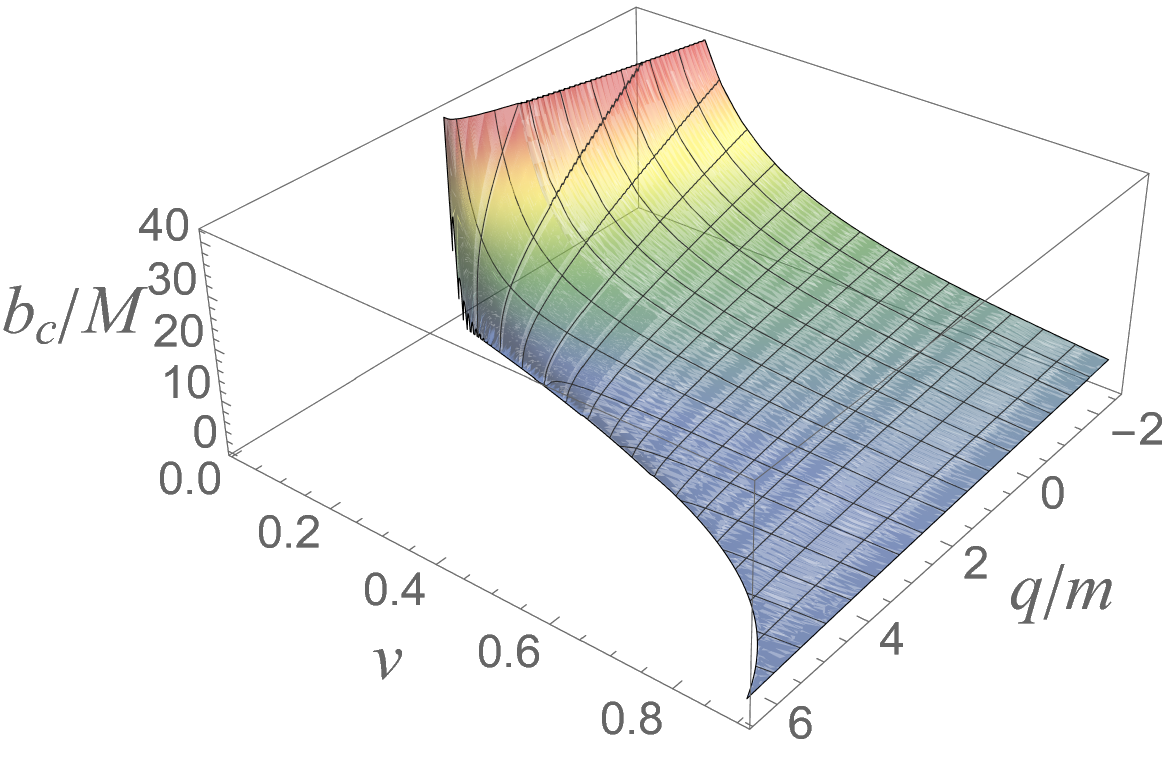}\\
	(b)
	\caption{(a) and (b) are respectively $r_{0c}$ and $b_c$ as functions of $q/m$ and $v$ for $Q/M=1/2$.  \label{fig:rcbc}}
\end{figure}

The $r_{0c}$ solved from Eq. \eqref{36} is plotted in Fig. \ref{fig:rcbc} (a) as a function of $q/m$ and $v$ for a typical $Q/M$. It is seen that for any fixed $v\leq 1$ and $0<Q/M\leq 1$, as $q$ increases to some positive maximal value $q_{\mathrm{max}}$, $r_{0c}$ decreases monotonically if $v$ is large or otherwise it could increase again when $q$ increases towards $q_{\mathrm{max}}$. Only for $q\leq q_{\mathrm{max}}$, a critical $r_{0c}$ is allowed. This implies that for any given $Q~(|Q|\leq M)$, as the repulsion between $q$ and $Q$ increases, the critical $r_{0c}$ may cease to exist. For cases with $q>q_{\mathrm{max}}$, signals incoming along any direction will not be captured. 
This $q_{\mathrm{max}}$ is actually determined by the requirement that even when $L=0=b$, there still exists a closest approach of radius and its value should be equal to $r_{0c}$. Solving $r_0$ from $L=0$ using Eq. \eqref{4}, the positiveness of $r_0$ yields the $q_{\mathrm{max}}$ for the RN spacetime to be
\be 
q_{\mathrm{max}}=\frac{m}{\sqrt{1-v^2}}\lb \frac{M}{Q}+v\sqrt{\frac{M^2}{Q^2}-1}\rb, \label{eq:r0cqbd} 
\ee
and at this boundary, the value or $r_{0c}$ is simply
\be 
r_{0c}(q=q_{\mathrm{max}},v)= M+\frac{\sqrt{M^2-Q^2}}{v}. \label{eq:r0catqbd}
\ee
Note that since $v<1$ this value is always larger than the RN BH exterior horizon radius $r_{\mathrm{H}}=M+\sqrt{M^2-Q^2}$.
The boundary \eqref{eq:r0cqbd} and its corresponding value of $r_{0c}$ is also plotted in Fig. \ref{fig:rcbc} (a) using the red curve. We also note that in the limit $v\to 0$, $q_\mathrm{max}$ approaches $mM/Q$ and $r_{0c}(q,v\to0)$ approaches an infinite value, except in the extremal RN case  $r_{0c}(q=q_\mathrm{max},Q=M,v)= M$. 

For any $r_{0c}$, there exists a corresponding critical value of the impact parameter, $b_c$.
For RN spacetime, using Eq. \eqref{13}, $b_c$ is found to be
\bea 
b_c&=&\frac{r_{0c}}{v}\lb \frac{1-v^2}{r_{0c}^2-2Mr_{0c}+Q^2}\rb^{1/2}\nn\\
&&\times \lsb \lb \frac{q}{m}Q-\frac{r_{0c}}{\sqrt{1-v^2}}\rb^2-\lb r_{0c}^2-2Mr_{0c}+Q^2\rb 
\rsb^{\frac{1}{2}}. \notag\\
&\label{eq:rnrc}
\eea
This $b_c$
is also important because it can be directly linked to the angular size $\theta_{\mathrm{sh}}$ of the BH shadow observed by the remote observer at radius $r_d$, through relation $ \theta_{\mathrm{sh}}\approx b_c/r_d  $ (see Eq. \eqref{46}) when $E$ and $r_d$ are large. Using Eq. \eqref{eq:rnrc}, the  $b_c$ corresponding to Fig. \ref{fig:rcbc} (a) is plotted in Fig. \ref{fig:rcbc} (b). It is seen that unlike $r_{0c}$, $b_c$ for any fixed $Q$ and $v$ decreases monotonically as $q$ increases, and reaches 0 at $q_{\mathrm{max}}$. While for its dependence on $v$, then as $v$ decreases to zero, $b_c$ increases monotonically to infinity. 

It is more interesting to study the effect of $Q$ on $b_c$ than that of $q/m$, because $q/m$ only affects the electric deflection while $Q$ affects both the electric and the gravitational deflections. Previous works have implied that in the weak field limit increasing $|Q|$ would decrease $b_c$ gravitationally for neutral signals (see Eq. (122) of Ref. \cite{Pang:2018jpm}), and the desired $b$ to reach the same observer will be increased if the signal is charged and $qQ<0$ (see by Eq. (3.3) of Ref. \cite{Xu:2021rld}). These mean that when $q$ is fixed and $qQ<0$, a nonzero $Q$ will cause a competition between its gravitational and electric effects on $b_c$, which is worthy to investigate. 

\begin{figure}[htp!]
	\includegraphics[width=0.45\textwidth]{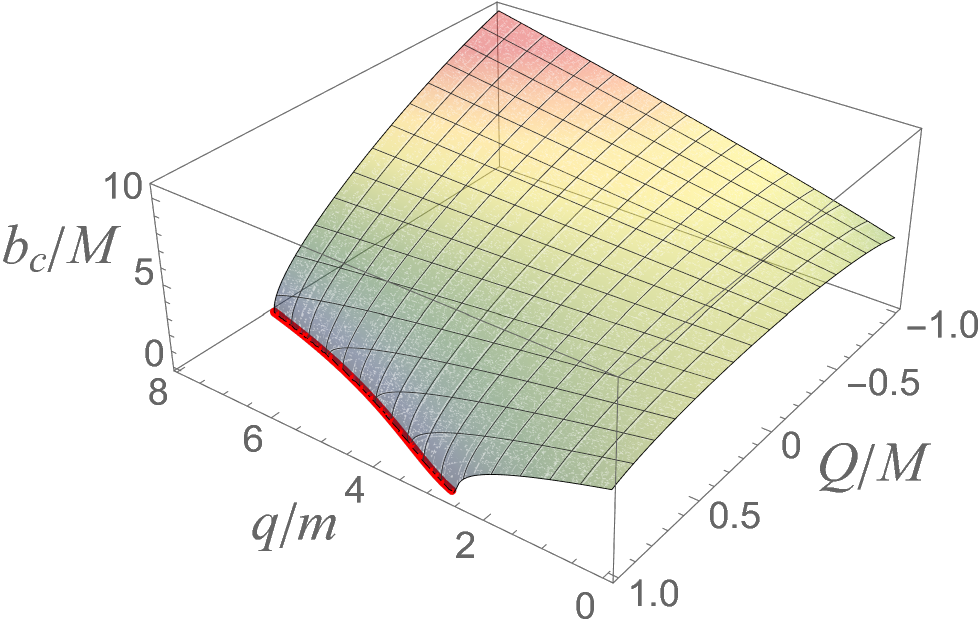}
	\caption{$b_{c}$ as a function of $Q/M$ and $q/m$ for $v=0.9c$. The red curve is the maximal $Q_{\mathrm{max}}$ given in Eq. \eqref{eq:cqmaxres} for a fixed $q$. \label{fig:rcbc2}}
\end{figure}

In Fig. \ref{fig:rcbc2} we plotted the $b_c$ as a function of $Q/M$ and $q/m$. It is seen that for fixed $q/m$, $b_c$ monotonically decreases as $Q$ increases, as long as it is positive so that $qQ>0$. For $q/m\gtrsim2.3$, when $Q$ increases to a certain point $Q_{\mathrm{max}}$, the $b_c$ decreases to zero. The critical points where $b_c$ approaches zero is also where $r_{0c}$ ceases to exist and therefore we can work out this $Q_{\mathrm{max}}$ directly from Eq. \eqref{eq:r0cqbd}. Replacing $q_{\mathrm{max}}$ by $q$ in this equation, we find 
\begin{align}
	Q_{\mathrm{max}}=\frac{Mm}{q}\frac{\sqrt{1-v^2}}{1-\sqrt{1-m^2/q^2}v}, \label{eq:cqmaxres}
\end{align}
which is shown as the red curve in Fig.  \ref{fig:rcbc2}. 
Using relation \eqref{495}, $b_c$ approaching zero corresponds to the point that the BH shadow angular size shrinks to zero. This will be verified in Fig. \ref{fig:thetash}. 
While for small $q/m$ ($0<q/m<2.3$), $b_c$ only decreases to a finite value when $Q$ reaches its extreme value of $Q_{\mathrm{ext}}=M$. 

On the other hand, for $Q<0$ so that $qQ<0$, we also see from Fig. \ref{fig:rcbc2} that $b_c$ also increases as $Q$ decreases by a small amount from zero (more apparent for large $q$). 
This suggests that for the chosen ranges of parameters $q$ and $v$, the electric effect on $b_c$ is stronger than that of the gravitational one for small negative $Q$, because we know that $Q$'s gravitational effect alone would decrease $b_c$ \cite{Xu:2021rld}. With $Q$ further decreases to more negative value, we see that for small $q/m$, $b_c$ decreases again, implying the gravitational effect exceeds the electric effect again. While for larger $q/m$, $b_c$ keeps increasing for the entire range of $Q$ from 0 to $-Q_{\mathrm{ext}}$. This will also be confirmed in Fig. \ref{fig:thetash} by ultra-high energy protons. The above features for small and large negative $Q$ are intuitively understandable after a quantitative comparison of the two effects on $\Delta\phi$ in the weak field limit. The gravitational and electric contributions  of $Q$ to $\Delta\phi$ in this limit are respectively about the size
\be -cQ^2/b^2~~\text{and}~~-c^\prime qQ/b,\label{eq:2contri}\ee
where $c$ and $c^\prime$ are some positive constants (see Eq. (4.4) of Ref. \cite{Xu:2021rld}). Consequently for a small but fixed $q$, when $Q$ is very small (or large) the gravitational effect on the deflection is always weaker (or stronger) than the electric one. 
Also from these contributions in Eq. \eqref{eq:2contri}, we see that if $q$ is very large, such as in the case of $q/m\gtrsim 2.3$ in Fig. \ref{fig:rcbc2}, the charge $Q$ might not be able to reach the desired value for the gravitational effect to overcome the electric one, before it reaches its extreme value $-M$.

Finally, it is also instructive to comment on the case when $q$ exceeds the $q_{\mathrm{max}}$ in Eq. \eqref{eq:r0cqbd}  for a fixed $Q$ or $Q$
exceeds the $Q_{\mathrm{max}}$ in Eq. \eqref{eq:cqmaxres} for a fixed $q$.
The vanishing of $r_{0c}$ or $b_c$ reaching zero do not mean that the signal will be able to travel to the very central region of the BH. Indeed in these cases the closet approach $r_0$ of the trajectory solvable from Eq. \eqref{4} still exists and can be shown to be well beyond the RN BH outer horizon. However, the critical behavior of the effective potential \eqref{eq:effpdef} detaching the right hand side of Eq. \eqref{eq:veffeqe} will not happen. A key noticeable feature following this is that the charged signal will always only experience a finite amount of angular deflection regardless how small is $b$. 

\subsection{Deflection angle $\Delta\phi$}

After obtaining $r_{0c}$,  we can continue to expand the RN metric and electric potential functions in Eq. \eqref{35} and obtain
\begin{subequations}\label{eq:mfexp}
	\begin{align}
		A(r)=     & \left(-\frac{2M}{r_{0c}}+\frac{Q^2}{r_{0c}^2}+1\right)-\frac{2\left(Q^2-Mr_{0c}\right)}{r_{0c}^3}(r-r_{0c}) \nn         \\
		& +\frac{ \left(3Q^2-2 M r_{0c}\right)}{r_{0c}^4}(r-r_{0c})^2 \nn\\
		& +\left(\frac{2 M}{r_{0c}^4}-\frac{4Q^2}{r_{0c}^5}\right)(r-r_{0c})^3+\mathcal{O} \left(r-r_{0c}\right)^4,  \label{39}      \\
		B(r)=     & \frac{1}{-\frac{2 M}{r_{0c}}+\frac{Q^2}{r_{0c}^2}+1}+\frac{2r_{0c}  \left(Q^2-M r_{0c}\right)}{\left(-2 Mr_{0c}+Q^2+r_{0c}^2\right)^2}(r-r_{0c})   \nn \\
		& +\frac{ \left(2M r_{0c}^3-3 Q^2 r_{0c}^2+Q^4\right)}{\left(-2 Mr_{0c}+Q^2+r_{0c}^2\right)^3}(r-r_{0c})^2   \nn             \\
		& -\frac{2\left(-M Q^4+M r_{0c}^4-2 Q^2 r_{0c}^3+2 Q^4r_{0c}\right)}{\left(-2 Mr_{0c}+Q^2+r_{0c}^2\right)^4}(r-r_{0c})^3 \nn   \\
		& +\mathcal{O} \left(r-r_{0c}\right)^4,      \label{40}         \\
		C(r)=     & r_{0c}^2+2 r_{0c}(r-r_{0c})+(r-r_{0c})^2,  \label{41}           \\
		A_{0}(r)= & -\frac{Q}{r_{0c}}+\frac{Q }{r_{0c}^2}(r-r_{0c})-\frac{Q}{r_{0c}^3}(r-r_{0c})^2  \nn        \\
		& +\frac{Q}{r_{0c}^4}(r-r_{0c})^3+\mathcal{O} \left(r-r_{0c}\right)^4. \label{42}
	\end{align}
\end{subequations}
Reading off the coefficients $a_{n}$, $b_{n}$, $c_{n}$ and $a_{0n}$ in Eq. \eqref{eq:metrica0exp} from the above, and substituting them into Eqs. \eqref{eq:yfirstfew}-\eqref{34}, the  $y_{-1,0},~y_{0,0}$ and higher order $y_{n,m}$'s can be obtained. Further substituting into Eq. \eqref{27b}, the perturbative $\Delta\phi$ in the RN spacetime in the SFL for charged signal is found immediately  
\begin{widetext}
	\begin{align}
		\Delta\phi_{R}\approx-\sqrt{2}y_{-1,0,R}\ln\left(1-\frac{b_{c}}{b}\right)+\sqrt{2}y_{-1,0,R}\ln\left(4\sqrt{\eta_s \eta_d}\right)+\sum_{i=s,d}\sum_{n=0}^{\infty}\frac{2y_{n,0,R}\eta_{i}^{\frac{n+1}{2}}}{2^{\left[\frac{n+1}{2}\right]+\frac{1}{2}}(n+1)}+\calco (a^1),\label{eq:dphiinrn}
	\end{align}
	where the coefficients are
	\begin{align}
		y_{-1,0,R}=&\frac{r_{0c}\sqrt{r_{0c}^2-2 M r_{0c}+Q^2}}{ \sqrt{2\left[r_{0c}^4 - 6 M r_{0c}^3 + 3(4 M^2 +  Q^2) r_{0c}^2 - 16 M Q^2 r_{0c}  + 6 Q^4  \right]}},\\
		y_{0,0,R}=&-\frac{r_{0c}\sqrt{r_{0c}^2-2 M r_{0c}+Q^2}}{\sqrt{2} 
			\left[r_{0c}^4 - 6 M r_{0c}^3 + 3(4 M^2 +  Q^2) r_{0c}^2 - 16 M Q^2 r_{0c}  + 6 Q^4  \right]^2}\notag\\
		&\times\left[r_{0c}^6-7 M
		r_{0c}^5+3\left(6 M^2+ Q^2\right) r_{0c}^4-5\left(4
		M^2+3  Q^2\right)M r_{0c}^3+2\left(16 M^2 + Q^2\right)Q^2 r_{0c}^2-18 M Q^4 r_{0c}+4 Q^6\right].
	\end{align}
	Higher order $y_{n,0}$ were also computed but they are numerically less important and too long to be shown here.
\end{widetext}

\begin{figure}
	\includegraphics[width=0.45\textwidth]{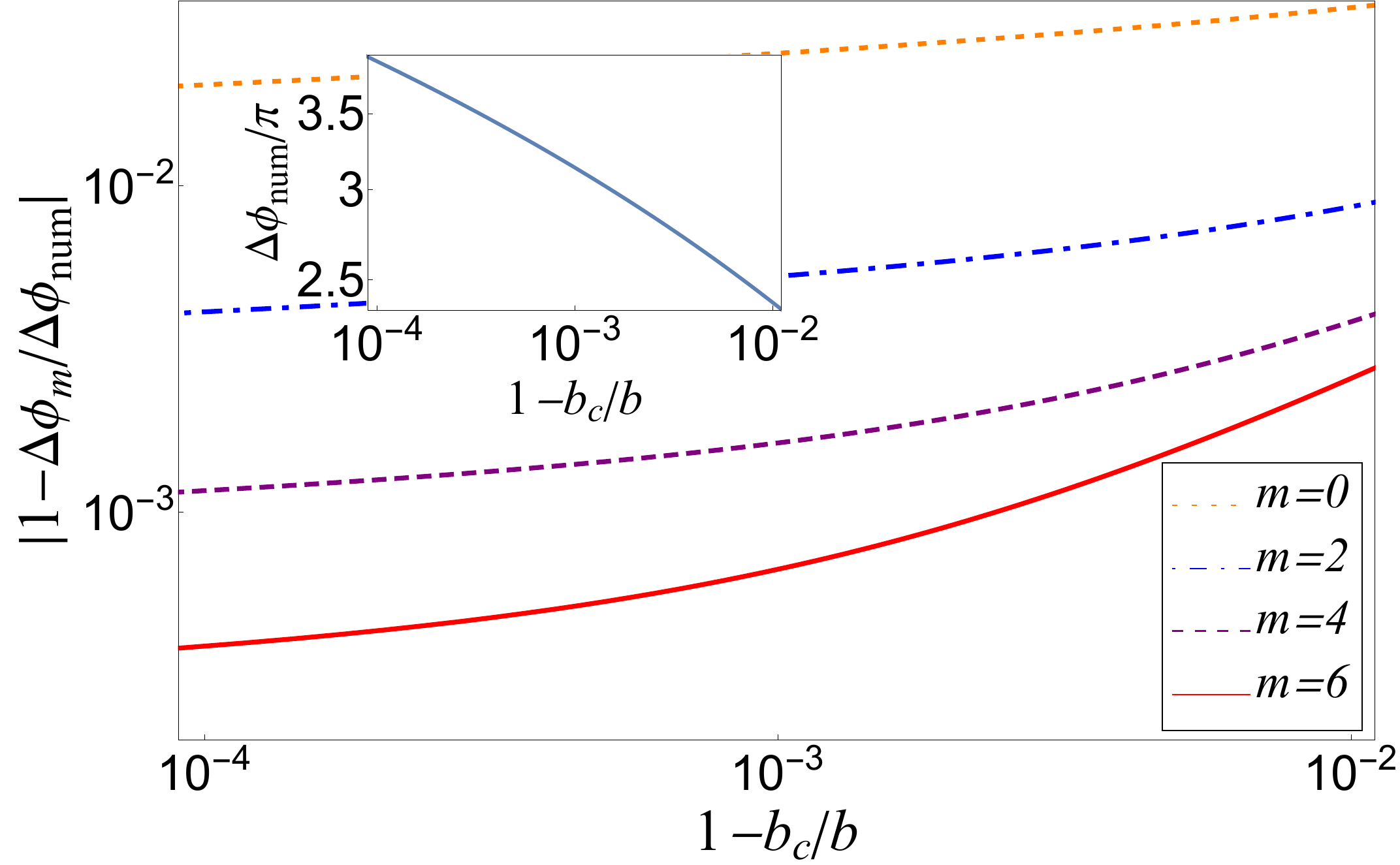}
	\includegraphics[width=0.45\textwidth]{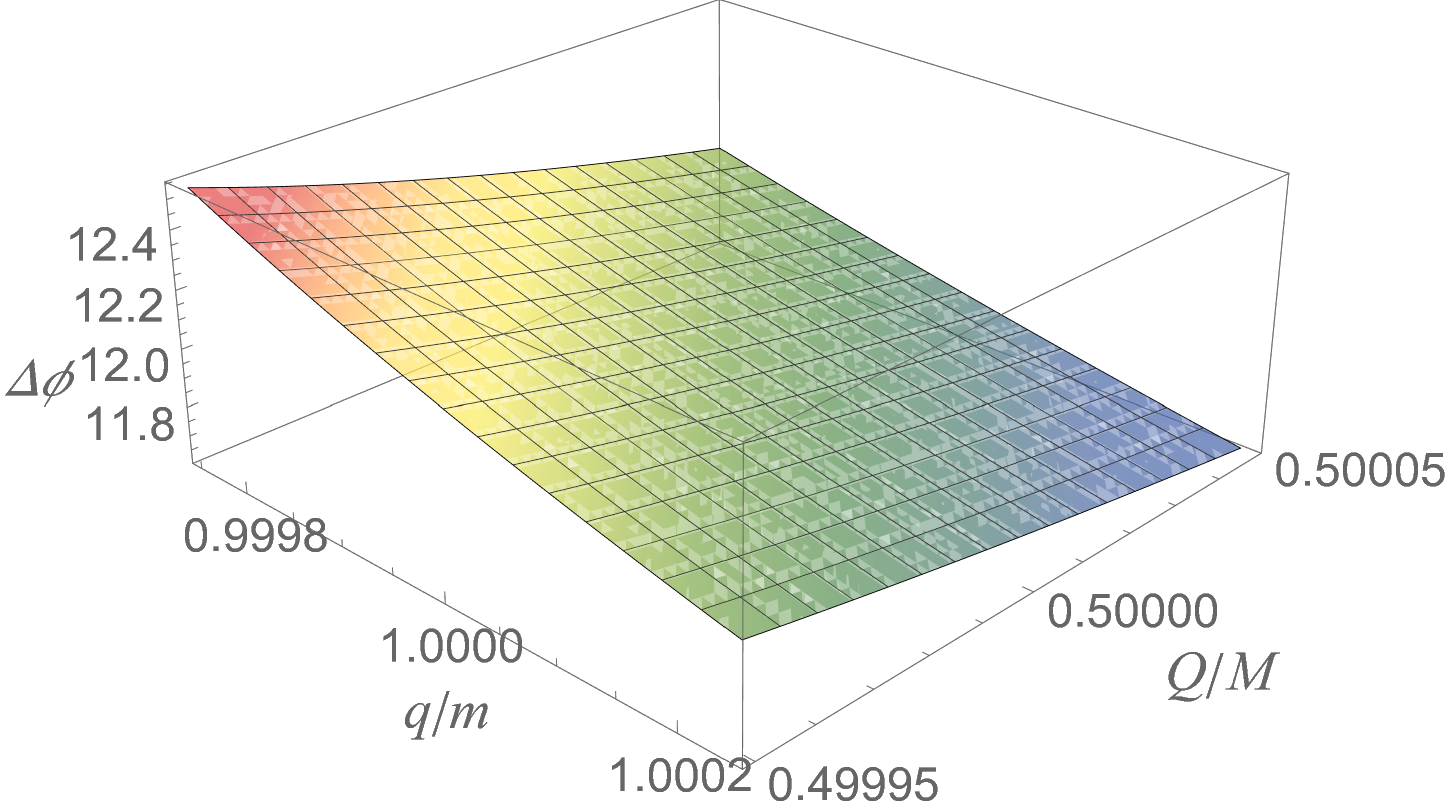}
	\caption{(a) Comparison between the truncated $\Delta\phi_m$ for $m=0,~2,~4,~6$ and the numerical integration result for $a=1-b_c/b$. (b) $\Delta\phi$ as a function of $Q/M$ and $q/m$. Other parameters are the same as in (a). \label{fig:dphicomp}}
\end{figure}

To check the validity of this $\Delta\phi$, we can define a truncated $\Delta\phi_m~(m=0,~2,~\cdots)$ by including only terms to order $m$ in the summation of $n$ in Eq. \eqref{eq:dphiinrn} and compare its value with a $\Delta\phi_{\mathrm{num}}$ obtained by numerically integrating Eq. \eqref{8}. In Fig. \ref{fig:dphicomp} (a) we chose some exemplary parameters $M=1,~Q=M/2,~q/m=1,~v=9/10$ and plot the $\Delta\phi_m$ and $\Delta\phi_{\mathrm{num}}$ as functions of $a=1-b_c/b$. 
We see that as the truncation order increases, the perturbative result converges to the numerical value in general, and more rapidly so for smaller $a$. This is in accord with the fact that as $a\to 0$, $\Delta\phi$ diverges as $\sim \ln a$ and the $a$-independent terms composing $D_0$ in Eq. \eqref{29} become less important to $\Delta\phi$. 

To study the effect of the charges $Q$ and $q$ and the electric interaction on the deflection in the SFL, in Fig. \ref{fig:dphicomp} (b) we plot $\Delta\phi$ using Eq. \eqref{eq:dphiinrn} for a fixed impact parameter. We chose a small parameter range of $Q\in(0.5\pm 0.0005)M$ and $q\in(1\pm0.002) m$ in order to make sure that for the fixed $b$, the parameter $a=1-b_c/b$ is still smaller than $1.4\times 10^{-4}$ so that the approximation \eqref{eq:dphiinrn} is always valid. 
It is seen that as $Q$ or $q$ increases, $\Delta\phi$ decreases monotonically in this range, suggesting that a stronger repulsion results in a smaller deflection. This actually is understandable from Fig. \ref{fig:rcbc2} that larger $Q$ and $q$ when $qQ>0$
result in a smaller $b_c$
and therefore the trajectory with fixed $b$ experiences weaker field and smaller deflection. 

\section{GL in the SFL AND BH SHADOW by charged signals \label{sec:glsflcs}}

To reveal the effect of the electric interaction on the GL in the SFL, we will have to solve the GL equation for this case. Formally, this equation and its solution process are the same as in the case of neutral particle which was studied in Ref. \cite{Jia:2020qzt}, except now all parameters or coefficients have to be updated to include the electric interaction. Therefore in this work, we will do a parallel analysis of the GL but concentrate on the electric effect this time. 

When one has a $\Delta\phi$ that takes the finite distance effect of the source and detector into account, establishing the GL is particularly simple. For a source located at $(r_s,\phi_{s})$ and detector at $(r_d,\phi_{d})~(0\leq \phi_d<\phi_s<2\pi)$ (see Fig. \ref{fig:glillus}), the change of the angular coordinate $\Delta\phi$ satisfies 
\begin{equation}\label{43}
	\Delta\phi(r_s,r_d,b)-2|n|\pi=\pi(1+\mathrm{sign}(n))-\mathrm{sign}(n)(\phi_{s}-\phi_{d}),
\end{equation}
where integer $|n|=1,~2,~\cdots$ is the looping number of the trajectory around the center and $n>0$ and $n<0$ corresponds to the anti-clockwise and the clockwise looping directions respectively.
Substituting Eq. \eqref{27b} for $\Delta\phi$ into Eq. \eqref{43}, it becomes
\begin{equation}\label{44}
	-C_{0}\ln \lb 1-\frac{b_c}{b}\rb +D_{0}-2n\pi=\pi(1+\mathrm{sign}(n))-\mathrm{sign}(n)\Delta\phi_{sd},
\end{equation}
where we have denoted $\Delta\phi_{sd}\equiv \phi_{s}-\phi_{d}$.
From this, we can easily solve $b$ that allows the signal to reach the detector in terms of other quantities
\begin{align}
	b_{n}= & \frac{b_c}{1-\exp \left\{\frac{-\left[2|n|+1+\mathrm{sign}(n)\right]\pi+D_{0}+\mathrm{sign}(n)\Delta\phi_{sd}}{C_{0}}\right\}},\nn\\
	&~~|n|=1,~2,~\cdots.  \label{45} 
\end{align}
Formally this is the same as Eq. (38) of Ref. \cite{Jia:2020qzt}, however the parameters $b_c,~C_0$ and $D_0$ are now given by the updated values in Eqs. \eqref{13}, \eqref{28} and \eqref{29}. 

\subsection{The relativistic image and BH shadow locations}

Corresponding to this series of $b_{n}$ are two series of images in the SFL, one series on each side of the lens, and their apparent angles are given by the following formula \cite{Xu:2021rld}
\begin{align}\label{46}
	\theta_n=&\arcsin\left[b_{n}\cdot \frac{\sqrt{E^{2}-m^{2}}}{\sqrt{\left(E+qA_{0}(r_d)\right)^{2}-m^{2}A(r_d)}}\sqrt{\frac{A_(r_d)}{C(r_d)}}\right],\nn\\
	&~|n|=1,~2,~\cdots .
\end{align}
Substituting Eqs. \eqref{45} and \eqref{13}, this becomes
\begin{align}
	\theta_n=&\arcsin\left[\frac{1}{1-\exp \left\{\frac{-\left[2|n|+1+\mathrm{sign}(n)\right]\pi+D_{0}+\mathrm{sign}(n)\Delta\phi_{sd}}{C_{0}}\right\}}\right. \nn \\
	&\times\left.\sqrt{\frac{\left(E+qA_{0}(r_{0c})\right)^{2}-m^{2}A(r_{0c})}{\left(E+qA_{0}(r_d)\right)^{2}-m^{2}A(r_d)}}\sqrt{\frac{A(r_d)C(r_{0c})}{A(r_{0c})C(r_d)}}\right] .\label{48}
\end{align}
Clearly if all other parameters are fixed but $|n|$ increases, $\theta_n$ will decrease monotonically. At $|n|\to\infty$, this yields the angular size $\theta_{\mathrm{sh}}$ of the BH shadow formed by charged signals. That is,
\begin{align}
	&\theta_{\mathrm{sh}}=\theta_{\pm\infty}\nn\\
	&=\arcsin\left[\sqrt{\frac{\left(E+qA_{0}(r_{0c})\right)^{2}-m^{2}A(r_{0c})}{\left(E+qA_{0}(r_d)\right)^{2}-m^{2}A(r_d)}}\sqrt{\frac{A(r_d)C(r_{0c})}{A(r_{0c})C(r_d)}}\right].   \label{495}
\end{align}
Clearly, the effect of first term inside the arcsin function in Eq. \eqref{48}, including parameters $\mathrm{sign}(n)$ and $\Delta_{sd}$, becomes irrelevant to the shadow size. 

\begin{figure}[htp!]
	\includegraphics[width=0.45\textwidth]{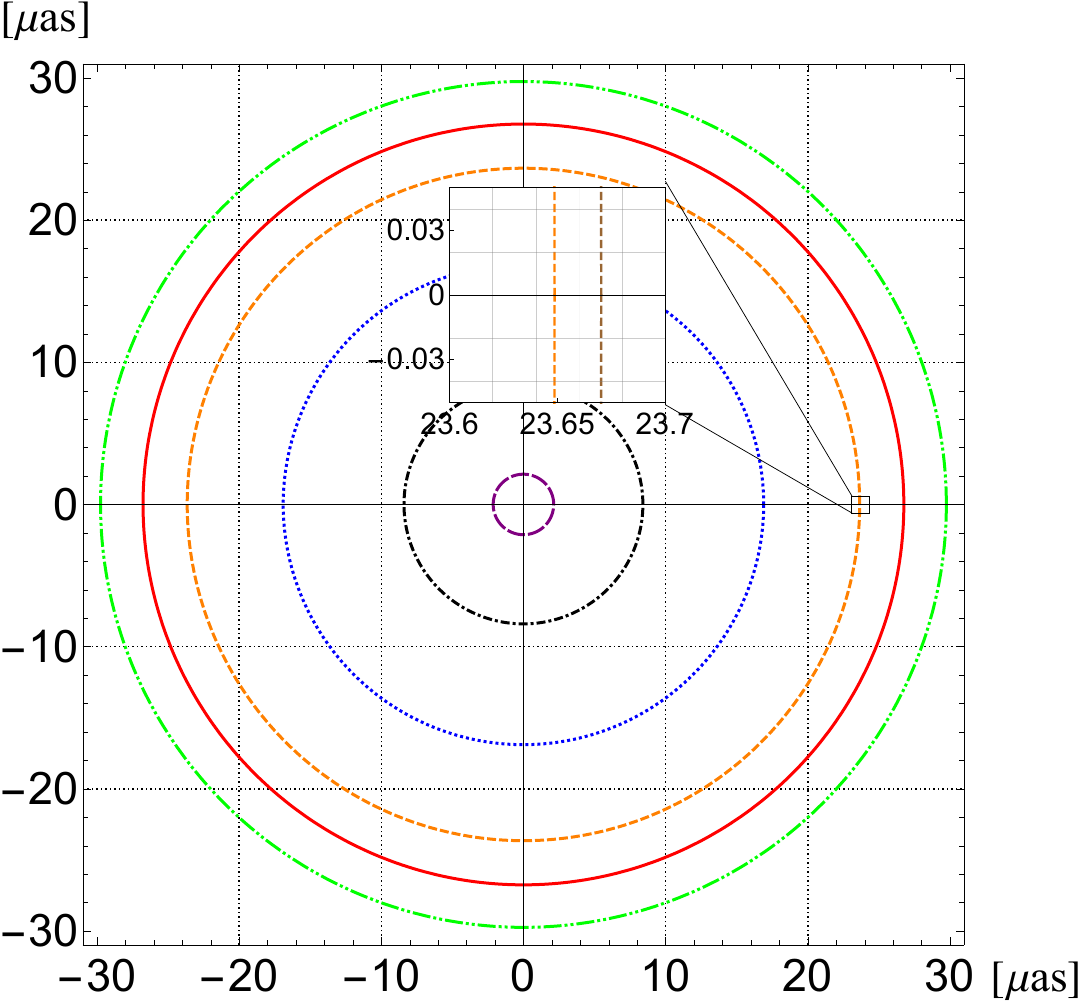}\\
	(a)\\
	\includegraphics[width=0.45\textwidth]{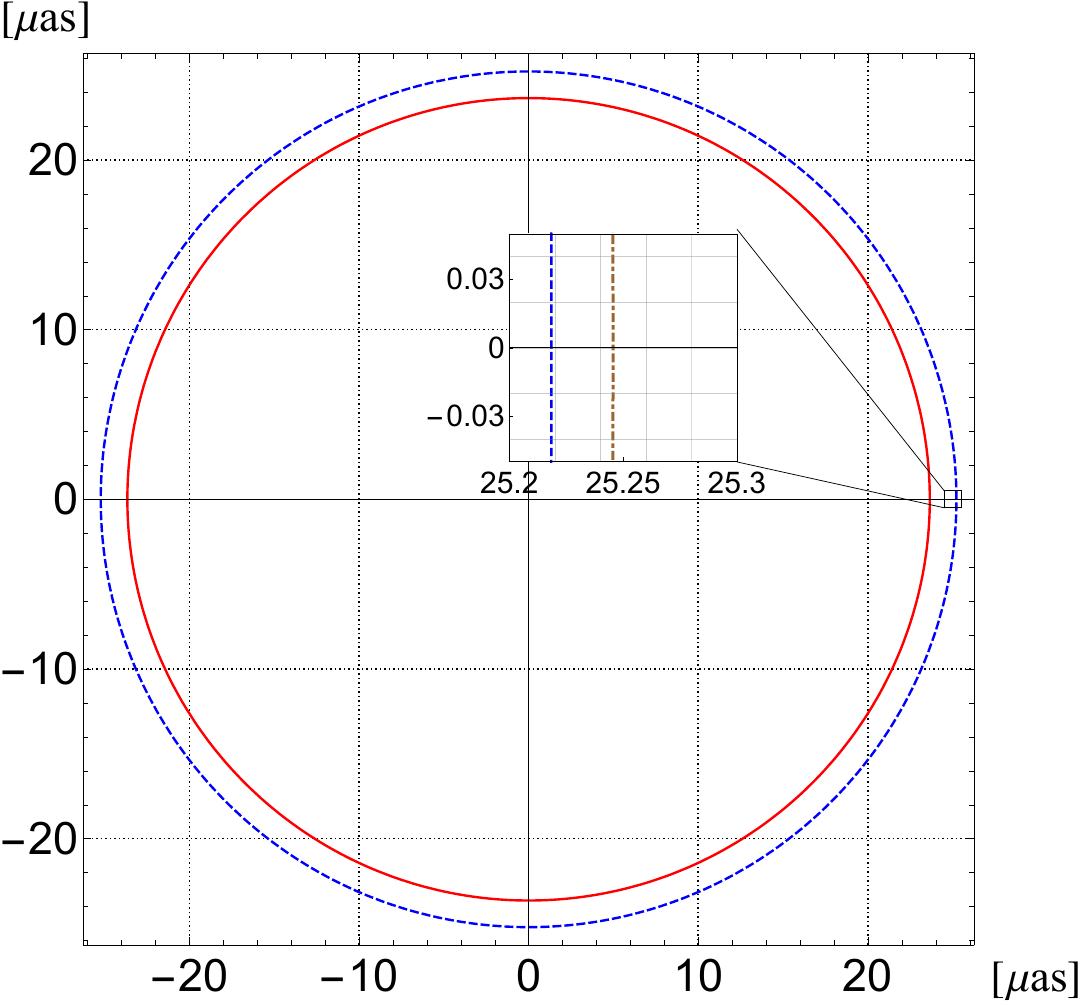}\\
	(b)
	\caption{The Einstein ring $\theta_{1}(\Delta\phi_{sd}=\pi)$ in the SFL using Eq. \eqref{48} and BH shadow $\theta_{\mathrm{sh}}$ using Eq. \eqref{495} for RN spacetime: (a) Formed by proton with energy $10^{19}$ [eV] for several $Q$. From outer to inner: $Q=-10^3Q_{\mathrm{mag}},~ 0,~10^3Q_{\mathrm{mag}},~3\times10^3Q_{\mathrm{mag}},~5\times10^3Q_{\mathrm{mag}},~5.8\times10^3Q_{\mathrm{mag}}$ respectively. (b) Formed by protons with energy $10^{19}$ [eV] and heavier nuclei with the same velocity  and $Q=10^{3}Q_{\mathrm{mag}}$. From inner to outer: proton and heavier nuclei with half $q/m$ of proton. Each line actually contains two separate lines: the outer $\theta_{1}$ and inner $\theta_{\mathrm{sh}}$ respectively (see the zoom-in).   \label{fig:thetash}}
\end{figure}

To see the effect of the electric interaction on $\theta_n$ and $\theta_{\mathrm{sh}}$, in Fig. \ref{fig:thetash} we plot them for the SgrA* SMBH by assuming it is a RN BH and the shadow is formed by cosmic protons. We take the proton energy to be $10^{19}$ [eV] and the SgrA* SMBH mass  $M=4.30\times10^{6}M_{\odot}$ and source/detector distance $r_{s}=r_{d}=8.28$[kpc] to be the distance to us \cite{GRAVITY:2021xju}. For the BH charge, we use the magnetically induced charge \begin{align}
	Q_{\mathrm{mag}}
	\approx 1.46\times 10^2\left(\frac{M}{M_\odot}\right)^2\frac{B_\mathrm{mag}}{10~[\text{G}]}~[\text{C}]
\end{align}
as the rough unit \cite{Xu:2021rld}. Here we take the typical order of $B_{\mathrm{mag}}\approx$10 [G]. We plot the first relativistic image $\theta_1$ and then the BH shadow $\theta_{\mathrm{sh}}$ for several $Q$. We chose $\Delta\phi_{sd}=\pi$ so that the source, lens and detector are aligned. From Fig. \ref{fig:thetash} (a) it is clear that for our fixed $q$, as $Q$ increases positively from zero (the red solid curve), the shadow size decreases monotonically, until $Q$ reaches $Q_{\mathrm{max}}$ at which point $b_c$ and consequently $\theta_{\mathrm{sh}}$ becomes zero. This corresponds to the large $q$ case in Fig. \ref{fig:rcbc2}, for which a maximal $Q$ exists. For the chosen parameter values  ($v,~q/m,~M$ etc.) in this figure, we can actually work out the critical $Q_{\mathrm{max}}=5862Q_{\mathrm{mag}}$ from Eq. \eqref{eq:cqmaxres}. 
Also, in this case, one can see from the green dash-dotted curve that for negative $Q$, the shadow size will increase, which again is in accord with Fig. \ref{fig:rcbc2}.  
Because of the same electrical repulsion effect, we see from Fig. \ref{fig:thetash} (b) that if charge $Q$ of the BH is fixed to a positive value, decreasing $q/m$ to its value of typical heavier nuclei, i.e., half its value for proton, will increase the shadow size. 
From both zoom-in's in Fig. \ref{fig:thetash} we see that all the relativistic images are very packed, and close to the BH shadow location. This is a general feature of all such lensed images in the SFL. 

\subsection{Magnification}

The magnification of the images are defined as
\begin{equation}\label{50}
	\mu_n=\left|\frac{\theta_n}{\beta} \frac{\mathrm{d}\theta_n}{\mathrm{d}\beta} \right|,
\end{equation}
where $\beta$ is the angle of source against the lens-detector axis.
To carry out the differentiation in above, we have to link $\beta$ to quantities in Eq. \eqref{48}. This can be achieved by considering the following geometrical relation 
\begin{equation}\label{51}
	\left[r_d-r_s\cos\Delta\phi_{sd}\right]\tan\beta=r_s\sin\Delta\phi_{sd}.
\end{equation}
Substituting Eq.\eqref{48} and Eq.\eqref{51} into Eq.\eqref{50}, we have 
\begin{align}
	\mu_n=&\left|\frac{\theta_n}{\beta} \frac{\mathrm{d}\theta_n}{\mathrm{d}\beta} \right|=\left|\frac{\theta_n}{\beta}\frac{\mathrm{d}\theta_n}{\mathrm{d}\Delta\phi_{sd}}\frac{\mathrm{d}\Delta\phi_{sd}}{\mathrm{d}\beta}\right| \nn \\
	\approx &\frac{1}{C_{0}}\frac{b_c^{2}}{r_sr_d^{3}}\frac{\left[r_d^{2}+r_s^{2}-2r_dr_s\cos\Delta\phi_{sd}\right]}{\left|(\cos\Delta\phi_{sd}-\frac{r_s}{r_d})\arctan\frac{\sin\Delta\phi_{sd}}{\cos\Delta\phi_{sd}-r_d/r_s}\right|}\nn\\
	&\times\exp {\frac{-(2|n|+1+\mathrm{sign}(n))\pi+D_{0}+\mathrm{sign}(n)\Delta\phi_{sd}}{C_{0}}}. \label{52}
\end{align}
Formally, this agrees with Eq. (44) of Ref. \cite{Jia:2020qzt} (after a typo corrected there), but again the coefficients $C_0,~D_0$ and $b_c$ are the ones containing the electric interaction.

\begin{figure}[htp!]
	\includegraphics[width=0.45\textwidth]{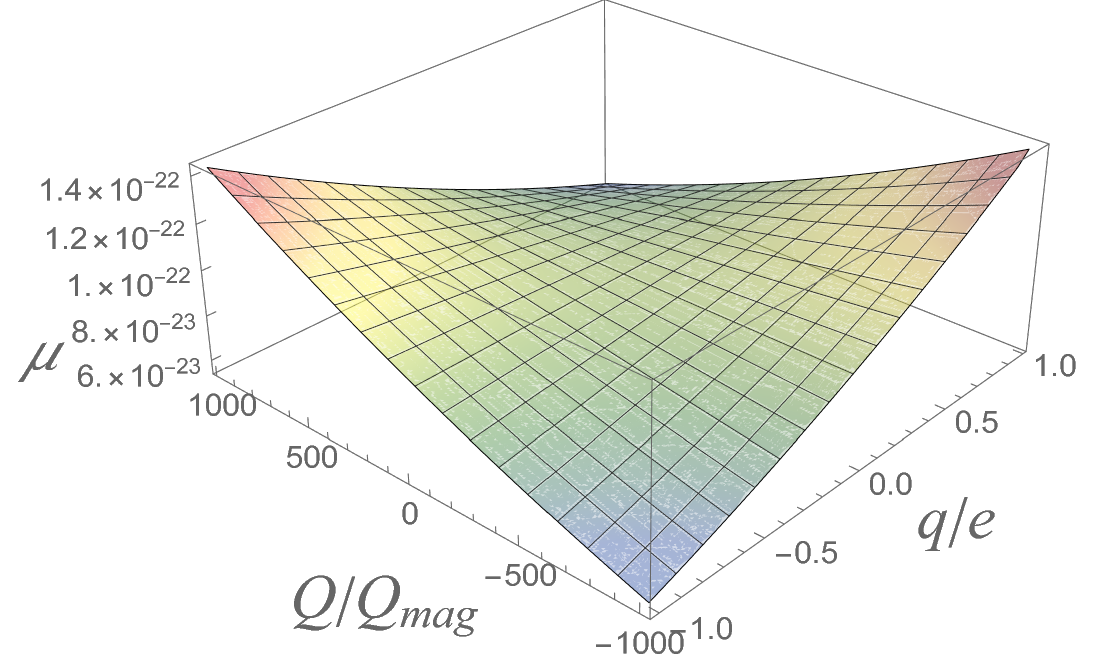}
	\caption{The magnification $\mu_n$ using Eq. \eqref{52} for $n=1,~\Delta\phi_{sd}=5\pi/6$. $q/e$ is the ratio of particle charge to elementary charge from $-1$ to $1$. Other parameters are the same as in Fig. \ref{fig:thetash}.  \label{fig:mag3d}}
\end{figure}

To see the electric effect on $\mu_n$ more clearly, in Fig. \ref{fig:mag3d} we have plotted the magnification as a function of $Q$ and $q$. 
It is seen that comparing to neutral particles, the magnification receives a correction that is also determined by $\mathrm{sign}(qQ)$. When $\mathrm{sign}(qQ)$ is positive (or negative), the magnification is decreased (or increased). Therefore qualitatively the effect of the electric interaction to the magnification here in the SFL is the same as that in the weak field limit \cite{Xu:2021rld}.

\subsection{Time delay}

With the impact parameter $b_n$ of the $n$-th images known in Eq. \eqref{45}, we can find the total travel time for the $n$-th trajectory using Eq. \eqref{27bp5}
\begin{align}
	\Delta t_n= -C'_{0}\ln \lb 1-\frac{b_{c}}{b_{n}}\rb +D'_{0}+\mathcal{O}(a^1). \label{eq:tnexp}
\end{align}
Using this, we can obtain the more observationally interesting quantity, the time delay $\Delta^2 t_{n,m}$ between the $n$-th and $m$-th images
\begin{align}
	&\Delta^2 t_{n,m}
	\equiv  \Delta t_{n}-\Delta t_{m}\nn\\
	=&\frac{2\pi C'_{0}}{C_{0}}\left[\left(|n|-|m|\right)+\frac{\left[\mathrm{sign}(n)-\mathrm{sign}(m)\right]\left(\pi-\Delta\phi_{sd}\right)}{2\pi}\right]. \label{eq:dt2mn}
\end{align}
where Eq. \eqref{45} for $b_n$ was substituted.

We now show that similar to the case of neutral particles  \cite{Liu:2021ckg}, Eq. \eqref{eq:dt2mn} also has a simple interpretation as the amount of local time to move around one full circle of the PS times the redshift factor from the PS to the observer, and then times the difference of the looping numbers of the two trajectories. For the first factor of Eq. \eqref{eq:dt2mn}, using Eqs. \eqref{29}, \eqref{C0'}, \eqref{32a} and \eqref{eq:zm10}, it becomes 
\be 
\frac{2\pi C'_{0}}{C_{0}}=\frac{2\pi\sqrt{c_{0}}}{\sqrt{1-\left(\frac{m}{E+a_{00}q}\right)^{2}a_{0}}}\frac{1}{\sqrt{a_{0}}}.\label{eq:firstfactor}
\ee
Now for a SSS spacetime, it is always possible to scale the metric function $C(r)=r^2$ and therefore its expansion at $r_{0c}$ yields $c_0=r_{0c}^2$. Consequently the numerator $2\pi \sqrt{c_0}$ of Eq. \eqref{eq:firstfactor} is actually the circumference of the PS. For the first denominator of this equation, we can show that it is nothing but the local speed $v_l$ of the particle's motion around the PS. Since in the SFL the signal circulates the PS, $v_l$ is given by
\be v_l=\frac{r_{0c}\dot{\phi}}{\gamma_{sr}}, \label{eq:vldef}
\ee 
where $\gamma_{sr}=1/\sqrt{1-v_l^2}$ is the gamma factor due to special relativity. Substituting Eqs. \eqref{3b}, \eqref{5a}, \eqref{13}, \eqref{eq:metrica0exp} and $c_0=r_{0c}^2$ into Eq. \eqref{eq:vldef} and after some simple algebra, one can solve $v_l$ as
\be 
v_l=\sqrt{1-\left(\frac{m}{E+a_{00}q}\right)^{2}a_{0}},\ee
which is exactly the first denominator of Eq. \eqref{eq:firstfactor}. Finally, the  factor $1/\sqrt{a_0}=1/\sqrt{A(r_{0c})}$ in Eq. \eqref{eq:firstfactor} is the redshift factor from the PS to an asymptotic observer in an SSS spacetime described by the metric \eqref{1}. 
Combining these three parts, therefore the first factor of Eq. \eqref{eq:dt2mn} is the time observed by the asymptotic observer for the signal to loop one circle of the PS.
The second factor of Eq. \eqref{eq:dt2mn} is the difference between looping number of the $n$-th and $m$-th trajectories.
Putting these factors together,
we obtain the claimed interpretation of the formula \eqref{eq:dt2mn}.
That is,
\begin{align}
	&\Delta^2 t_{n,m}=\frac{2\pi r_{0c}}{\sqrt{1-\left(\frac{m}{E+a_{00}q}\right)^{2}a_{0}}}\frac{1}{\sqrt{a_{0}}}\nn\\
	&\times\left[\left(|n|-|m|\right)+\frac{\left[\mathrm{sign}(n)-\mathrm{sign}(m)\right]\left(\pi-\Delta\phi_{sd}\right)}{2\pi}\right]. \label{eq:tdsim}
\end{align}

\begin{figure}[htp!]
	\includegraphics[width=0.45\textwidth]{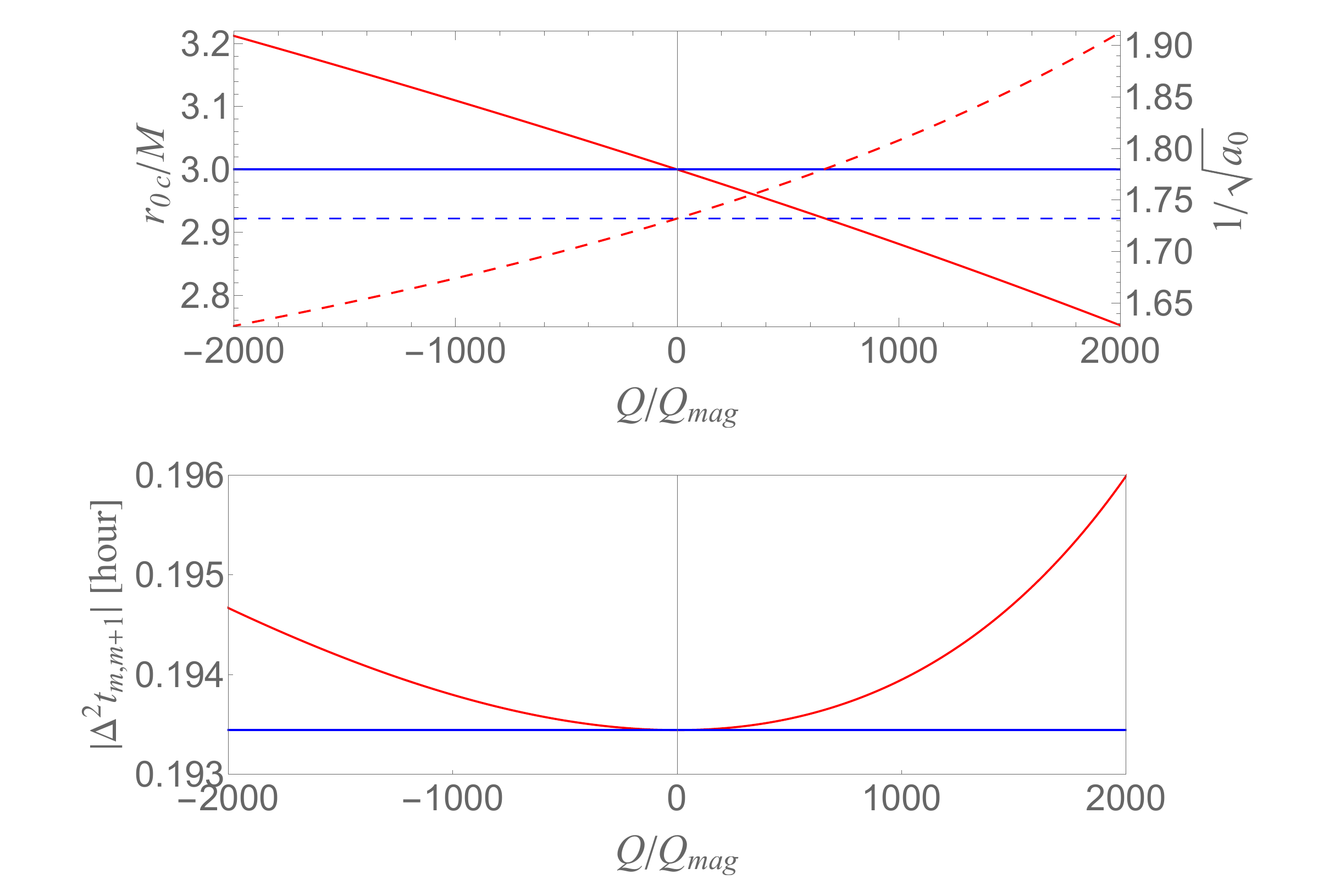}\\
	(a)\\
	\includegraphics[width=0.45\textwidth]{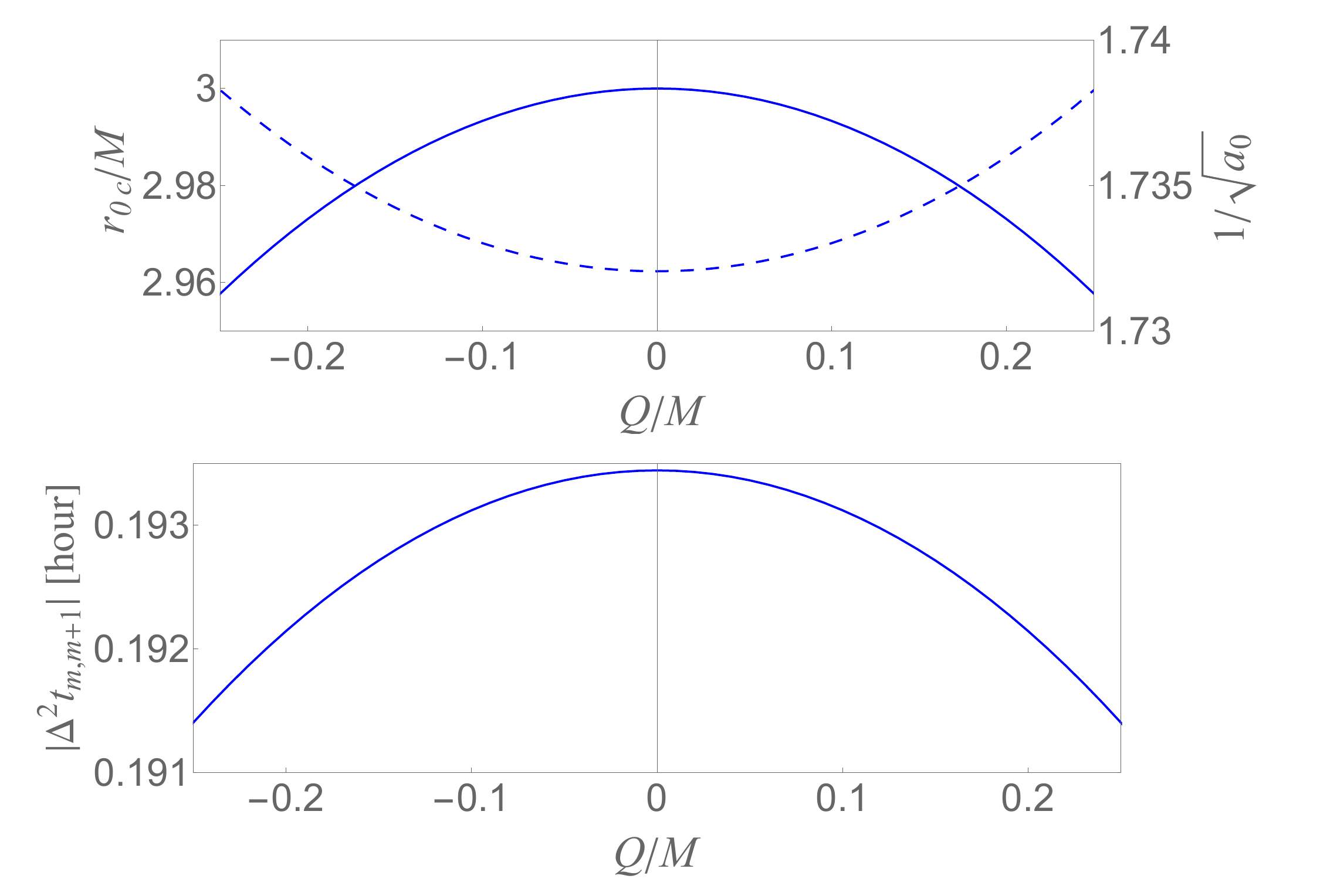}\\
	(b)\\
	\caption{(a): Upper: the PS radius (solid curves) and the redshift factor (dashed curves) as functions of $Q$ for protons (red curves) with energy $10^{19}$ [eV] and neutral signal (blue curves) with Sgr A* as the lens. Lower: the corresponding time delay $|\Delta^2 t_{m,m+1}|$. (b): the case for neutral signal for a much larger range of $Q$.  \label{fig:tdoncq}}
\end{figure}

In the above time delay, usually  $E,~m$ and $q$ of the signal are measurable. And if we assume that the spacetime is RN type, then it is seen from Eqs. \eqref{36}, \eqref{39} and \eqref{42} that $r_{0c},~a_0$ and $a_{00}$ will be fixed once $M$ and $Q$ are known. Since mass $M$ of a lens can often be known using other astronomical means, from Eq. \eqref{eq:tdsim} we immediately see that the charge $Q$ can be constrained by the measurement of $\Delta^2 t_{n,m}$.
In Fig. \ref{fig:tdoncq}, we plot the PS radius $r_{0c}$, the redshift factor $1/\sqrt{a_0}$ and the time delay $\Delta^2 t_{m+1,m}$ 
as functions of $Q$ by again assuming the Sgr A* SMBH is a RN one. Note that $\Delta^2 t_{m+1,m}$ actually does not depend on the value of $\Delta \phi_{sd}$ or $m$ as long as $m\geq 1$ or $m\leq -2$. It is seen from the lower subplot of Fig. \ref{fig:tdoncq} (a) that
for charged signal with $q>0$, the time delay increases when $Q$ deviates from zero. This is actually a combine effect of a decreasing PS radius and increasing redshift factor (see the upper subplot of Fig. \ref{fig:tdoncq} (a)). While for neutral signals, although all quantities seem flat in Fig. \ref{fig:tdoncq} (a), they are not truly invariant. They appear quite flat because in this case there is no electric interaction and therefore $Q$ only affect them gravitationally and then it takes a much larger $Q$ to make a comparable difference. As one can see from Fig. \ref{fig:tdoncq} (b), only when $Q$ reaches about $0.25M\approx 7.6\times 10^{10}Q_{\mathrm{mag}}$ , the time delay variation can reach the same order as the charged case in Fig. \ref{fig:tdoncq} (a). This comparison suggests that although the time delay of the neutral signal can also be used to constrain the lens charge as suggested in Ref. \cite{Liu:2021ckg}, that of the charged signal is more sensitive when $Q$ is small, which is generally expected by many astronomers. 

\section{Conclusions \label{sec:disc}}

In this work we studied the deflection, GL and time delay of charged signal in a charged spacetime in the SFL using a perturbative method. The effect of the extra electric interaction was carefully analyzed. It is found that the deflection $\Delta\phi$ still has a weak logarithmic divergence 
\begin{align}
	\Delta\phi =\sum_{n=0}^\infty \lsb -C_n\ln \lb 1-\frac{b_c}{b}\rb +D_n\rsb \lb 1-\frac{b_c}{b}\rb^n
\end{align}
as the trajectory approaches the PS. Perturbatively, a small electric repulsion (or attraction) will decrease (or increase) the critical impact parameter and therefore decreases (or increases) the deflection. For any fixed $Q$, in general there is a maximal $q_{\mathrm{max}}$ given in Eq. \eqref{eq:r0cqbd} ($qQ>0$) beyond which the PS ceases to exist and $b_c$ shrinks to zero. While for fixed $q$, the electric effect of $Q$ on $b_c$ will be weaker (or stronger) than its gravitational one when $Q$ is small (or large). 

For the GL in the SFL, similar to the neutral signal case, there exists one series of very packed and weakly magnified images on each side of the lens. Their angular positions and magnifications are given by Eqs. \eqref{48} and \eqref{52}. The BH shadow with electric interaction is still given by a simple formula Eq. \eqref{495}.
The electric repulsion (or attraction) tends to decrease (or increase) the angular sizes of these images and that of the BH shadow.
For a fixed $q/m$ of the charged signal, the BH shadow size shrinks to zero when $Q$ is beyond $Q_{\mathrm{max}}$ ($qQ>0$). 

The time delay between two images given by Eq. \eqref{eq:tdsim} has an intuitive and yet quantitatively precise explanation, as the PS circumference dividing the local velocity and then multiplying the redshift factor from the PS to the observer and the difference of the loop numbers of the two trajectories. It is shown that comparing to neutral signals, the time delay of charged signal is much more sensitive to the spacetime charge $Q$ when it is small. 

We emphasize that most of the results in this work, except Sec. \ref{sec:dphirn}, are completely general to all SSS charged spacetimes. Therefore besides the RN one, it is equally simple to apply them to other kinds of charged spacetimes, such as the Gibbons-Maeda-Garfinkle-Horowitz-Strominger (GMGHS) \cite{Gibbons:1982ih,Gibbons:1987ps,Garfinkle:1990qj}
and charged Horndeski \cite{Cisterna:2014nua} spacetimes. We also expect that it is straightforward to extend the perturbative method in this work to the deflection and GL in the equatorial plane of stationary and axially symmetric spacetimes, with or without the electric interactions.

\begin{acknowledgements}
	The authors would like to thank Haotian Liu for helping creating illustration Fig. \ref{fig:glillus}. This work is partially supported by the NNSF China
	11504276.
	
\end{acknowledgements}

\end{document}